\documentclass[iop,numberedappendix,appendixfloats]{emulateapj}

\usepackage{epstopdf}
\usepackage{graphics}
\usepackage{natbib}

\newcommand{\msun}{\mbox{$\rm M_{\odot}$}}

\newcommand{\zfn}{\mbox{$\zeta_{\rm MS}$}}

\newcommand{\lsun}{\mbox{$L_{\odot}$}~}
\newcommand{\kms}{\mbox{km s$^{-1}$}}

\newcommand{\etal}[1]{{ et al.}~}
\def\kms{\ifmmode \hbox{km~s}^{-1}\else km~s$^{-1}$\fi}
\def\etal {{\it et al.}}
\def\deg      {{\ifmmode^\circ\else$^\circ$\fi} } 

\def\h2     {H$_2$}

\def\h2     {H$_2$}

\def\arcmin{\hbox{$^\prime$}}
\def\arcsec{\hbox{$^{\prime\prime}$}}

\begin{document}

\shorttitle{Evolution of ISM, SF, and Accretion at High-Redshift}
\shortauthors{Scoville \etal}

\title{Evolution of Interstellar Medium, Star Formation, and Accretion \\ at High Redshift}

 \author{ N. Scoville\altaffilmark{1}, N. Lee\altaffilmark{3}, P. Vanden Bout\altaffilmark{2},  T. Diaz-Santos\altaffilmark{18}, D. Sanders\altaffilmark{8}, B. Darvish\altaffilmark{1}, A. Bongiorno\altaffilmark{4},\\ C. M.  Casey\altaffilmark{5}, L. Murchikova\altaffilmark{1}, J. Koda\altaffilmark{6},  P. Capak\altaffilmark{7},  Catherine Vlahakis\altaffilmark{9}, O. Ilbert\altaffilmark{14}, K. Sheth\altaffilmark{10}, \\ K. Morokuma-Matsui\altaffilmark{11}, 
 R. J. Ivison\altaffilmark{16,17}, H. Aussel\altaffilmark{12}, C. Laigle\altaffilmark{13}, H. J. McCracken\altaffilmark{13},  L. Armus \altaffilmark{7}, \\ A. Pope\altaffilmark{15}, S. Toft\altaffilmark{3}, and D.Masters\altaffilmark{7}}

\altaffiltext{1}{California Institute of Technology, MC 249-17, 1200 East California Boulevard, Pasadena, CA 91125}
\altaffiltext{2}{National Radio Astronomy Observatory, 520 Edgemont Road, Charlottesville, VA 22901, USA}
\altaffiltext{3}{AD(Dark Cosmology Centre, Niels Bohr Institute, University of Copenhagen, Juliana Mariesvej 30, DK-2100 Copenhagen, Denmark)}
\altaffiltext{4}{INAF - Osservatorio Astronomico di Roma, Via di Frascati 33, I-00040 Monteporzio Catone, Rome, Italy}
\altaffiltext{5}{Department of Astronomy, The University of Texas at Austin, 2515 Speedway Blvd Stop C1400, Austin, TX 78712}
\altaffiltext{6}{Department of Physics and Astronomy, SUNY Stony Brook, Stony Brook, NY 11794-3800, USA}
\altaffiltext{7}{Spitzer Science Center, MS 314-6, California Institute of Technology, Pasadena, CA 91125}
\altaffiltext{8}{Institute for Astronomy, 2680 Woodlawn Dr., University of Hawaii, Honolulu, Hawaii, 96822}
\altaffiltext{9}{North American ALMA Science Center, National Radio Astronomy Observatory, 520 Edgemont Road, Charlottesville, VA 22901, USA}
\altaffiltext{10}{NASA Headquarters, 300 E Street SW, Washington DC 20546}
\altaffiltext{11}{Chile Observatory, National Astronomical Observatory of Japan, 2-21-1 Osawa, Mitaka, Tokyo 181Ð8588, Japan}
\altaffiltext{12}{AIM Unit\'e Mixte de Recherche CEA CNRS, Universit\'e Paris VII UMR n158, Paris, France}
\altaffiltext{13}{ CNRS, UMR 7095, Institut dÕAstrophysique de Paris, F- 75014, Paris, France}
\altaffiltext{14}{Laboratoire dÕAstrophysique de MarseilleÑLAM, UniversitŽ dÕAix-Marseille \& CNRS, UMR7326, 38 rue F. Joliot-Curie, F-13388 Marseille Cedex 13, France}
\altaffiltext{15}{Department of Astronomy, University of Massachusetts, Amherst, MA 01003}
\altaffiltext{16}{Institute for Astronomy, University of Edinburgh, Blackford Hill, Edinburgh EH9 3HJ, UK}
\altaffiltext{17}{European Southern Observatory, Karl-Schwarzschild-Strasse 2, D-85748 Garching bei Munchen, Germany}
\altaffiltext{18}{Nœcleo de Astronom'a de la Facultad de Ingenier'a, Universidad Diego Portales, Av. EjŽrcito Libertador 441, Santiago, Chile}

{}

\begin{abstract} 
ALMA observations of the long wavelength dust continuum are used to estimate the interstellar medium (ISM) masses in a sample of 708 galaxies  at z = 0.3 to 4.5 in the COSMOS field. The galaxy sample has known far-infrared luminosities and, hence, star formation rates (SFRs), and stellar masses (M$_{\rm *}$)  from the optical-infrared spectrum fitting. The galaxies sample SFRs from the main sequence (MS) to 50 times above the MS. The derived ISM masses are used to determine the dependence of gas mass on redshift, M$_{\rm *}$, and specific SFR (sSFR) relative to the MS.  The ISM masses increase approximately 0.63 power of the rate of increase in SFRs with redshift and the 0.32 power of the sSFR/sSFR$_MS$. The SF efficiencies also increase as the 0.36 power of the SFR redshift evolutionary and the 0.7 power of the elevation above the MS; thus the increased activities at early epochs are driven by both increased ISM masses and SF efficiency. Using the derived ISM mass function we estimate the accretion rates of gas required to maintain continuity of the MS evolution ($>100$ \msun yr$^{-1}$ at z $>$ 2.5).  Simple power-law dependences are similarly derived for the gas accretion rates.  We argue that the overall evolution of galaxies is driven by the rates of gas accretion. The cosmic evolution of total ISM mass is estimated and linked to the evolution of SF and AGN activity at early epochs.
\end{abstract} 


\section{Galaxy Evolution at High Redshift}

Galaxy evolution in the early universe is dominated by three processes: the conversion of interstellar gas into stars, the accretion of 
intergalactic gas to replenish the interstellar gas reservoir, and the merging of galaxies. The latter redistributes the 
gas within the galaxies, promotes starburst activity fuels active galactic nuclei (AGN) and transforms the stellar morphology from disk-like (rotation dominated) to ellipsoidal systems. In all of these processes, the interstellar gases play a determining role -- including in the outcome of galaxy mergers, since the 
gas is dissipative and becomes centrally concentrated, fueling starbursts and AGN. 

At present the gas supply and its evolution at high redshifts are only loosely constrained \citep[see the reviews -- ][]{sol05,car13}. Only $\lesssim$ 200 galaxies at z $>$1 have been observed in 
the CO lines (and most not in the CO (1-0) line  which has been calibrated as a mass tracer.) To properly chart and understand the evolution large samples are required -- probing multiple characteristics: 1) the variation with redshift or cosmic 
time, 2) the dependence on galaxy mass, and 3) the differences between the galaxies with 'normal' SF activity on the main sequence (MS) and the starbursts (SB). The latter 
constitute only $\sim$ 5\% of the population but have SFRs elevated to 2 - 100 times higher levels than the MS. The contribution of the SB galaxies to 
the total SF at z $<$ 2 is 8 - 14\% \citep{sar12}. Although the high-z galaxies above the MS will here be referred to as starbursts (SB), nearly all high-z SF galaxies would be classified as SB galaxies if they were at low redshift.

Properly constraining the evolution of the gas contents would require time-consuming CO line observations spanning and populating the full ranges of redshift, $M_{\rm *}$ and MS versus SB populations. In addition, translating the observed, excited state CO emission fluxes into reliable estimates of the gas contents remains a problem \citep{car13}.

As an alternative, we here develop a formulation for the high redshift galaxy evolution, using very extensive observations of the Rayleigh-Jeans (RJ) dust continuum emission, and applying the 
calibration of this technique, developed in \cite{sco16} (a brief summary is provided here in Section \ref{rj_dust} and Appendix \ref{dust_app}) (see also \citep{mag14,san14,gen15,sch16,ber16}). The calibration used here is based on observations of the RJ dust emission and CO (1-0) emission in low redshift galaxies; the technique provides
roughly a factor two accuracy in the derived ISM masses -- provided one restricts the galaxy samples to high stellar mass galaxies ($> 10^{10}$\msun), which are expected to have near-solar  
gas phase metallicity \citep[e.g.][]{erb06}. Such galaxies are expected to have gas-to-dust abundance ratios and dust properties similar to those galaxies in our calibration sample at low redshift \citep{dra07b}. 

The long wavelength dust emission is nearly always 
optically thin and is only mildly susceptible to `excitation' variations, since the RJ emission depends  linearly on the mass-weighted dust temperature, and there is no dependence 
on the volume density of the gas and dust. In fact, the mass-weighted ISM dust temperatures are likely to vary less than a factor 2 (Section \ref{rj_td_dust}, see also \citep{sco16}. We note also that in contrast to the CO, which typically constitutes just $\sim$0.1\% of the 
gas mass,  the dust is a 1\% mass tracer. CO is also susceptible to depletion by photo-dissociation in strong UV radiation fields, whereas the dust abundance is less sensitive to the ambient radiation field.

The galaxy sample used here has ALMA continuum observations in Band 6 (240 GHz) and Band 7 (345 GHz); they are all within the COSMOS survey field and thus have excellent ancillary data. The ALMA pointings are non-contiguous 
but their fields of view (FOVs), totaling 102.9 arcmin$^2$ include 708 galaxies measured at far-infrared wavelengths by Herschel. All of the 
Herschel sources within the ALMA FOVs are detected by ALMA, so this is a {\it complete} sampling of the IR-bright galaxies at z = 0.3 to 4.5. Given the positional prior from the COSMOS Herschel catalog, all sources are detected in the far-infrared; the sample therefore has reliable estimates of the dust-embedded
SFR activity. The dusty SF activity is in virtually all cases dominant (5-10 times) over the unobscured SF probed in the optical/UV. 

\medskip
The major questions we address are:
\begin{enumerate}
\item How does the gas content depend on the stellar mass of the galaxies?
\item How do these gas contents evolve with cosmic time, down to the present, where they are typically less than 10\% of the galactic mass?
\item In the starburst populations, is the prodigious SF activity driven by increased gas supply or increased efficiency for converting the existing gas into stars?
\item How rapidly is the ISM being depleted? The depletion timescale is characterized by the ratio $M_{\rm ISM}$/SFR, but to date this has not 
been measured in broad samples of galaxies due to the paucity of quantitative ISM measurements at high redshift.
\item If the ISM depletion times are as short as they appear to be (several $\times10^8$ yrs), yet the SF galaxy population persists for a much longer span of cosmic time ($\gtrsim5\times10^9$ yrs), 
then there must be large rates of accretion of new gas from the halo or intergalactic medium (IGM) to maintain the activity. At present there are virtually no observational constraints on these accretion rates, only theoretical
predictions \citep{dek13}.; so a major unknown is what are these accretion rates?
\item How does the efficiency of star formation from a given mass of ISM (the so-called gas-depletion timescale) vary with cosmic epoch, the stellar mass of the galaxy, 
and whether the galaxy is on the MS or undergoing a starburst?
\end{enumerate}

\subsection{Outline of this Work}

We first provide a brief background to our investigation and develop the logical framework for tracking the galaxy evolution in Section \ref{basis}. We  then describe the datasets used for this investigation and the galaxy sample selection (Section \ref{datasets}).
 A brief background on the use of the RJ dust emission to probe ISM masses is given in Section \ref{rj_dust}; a more complete description 
is provided in Appendix \ref{dust_app} and \cite{sco16}. 

In Section \ref{fits},
analytic fits are obtained for the dependence of the ISM masses and SFRs on galaxy properties and redshift. 
These expressions are then used to elucidate the evolution of the ISM contents and the efficiencies for star formation as a function of redshift, stellar mass and sSFR relative to the MS. 
Section \ref{relations} uses these results to determine the gas depletion timescales and gas mass fractions. Section \ref{accretion} provides estimates of the gas accretion rates required to 
maintain the SF in the galaxy population and Section \ref{sb_ms} provides a brief discussion of the differences between the starburst and the MS galaxies. The derived equations are summarized in Tables
\ref{equations1} and \ref{equations2}.

The overall cosmic evolution of the 
ISM contents and the gas mass fractions integrated over the galaxy population from $M_* = 10^{10} - 10^{12}$ \msun~ is presented in Section \ref{cosmic}. 
A check on the self-consistency of these relations is 
provided in Section \ref{check} and a comparison with previous 
results based on CO and dust measurements is in given in Section 
\ref{previous}. Section \ref{summary} provides a summary of the main 
results and their implications. 

\begin{figure*}[ht]
\epsscale{1.}  
\plotone{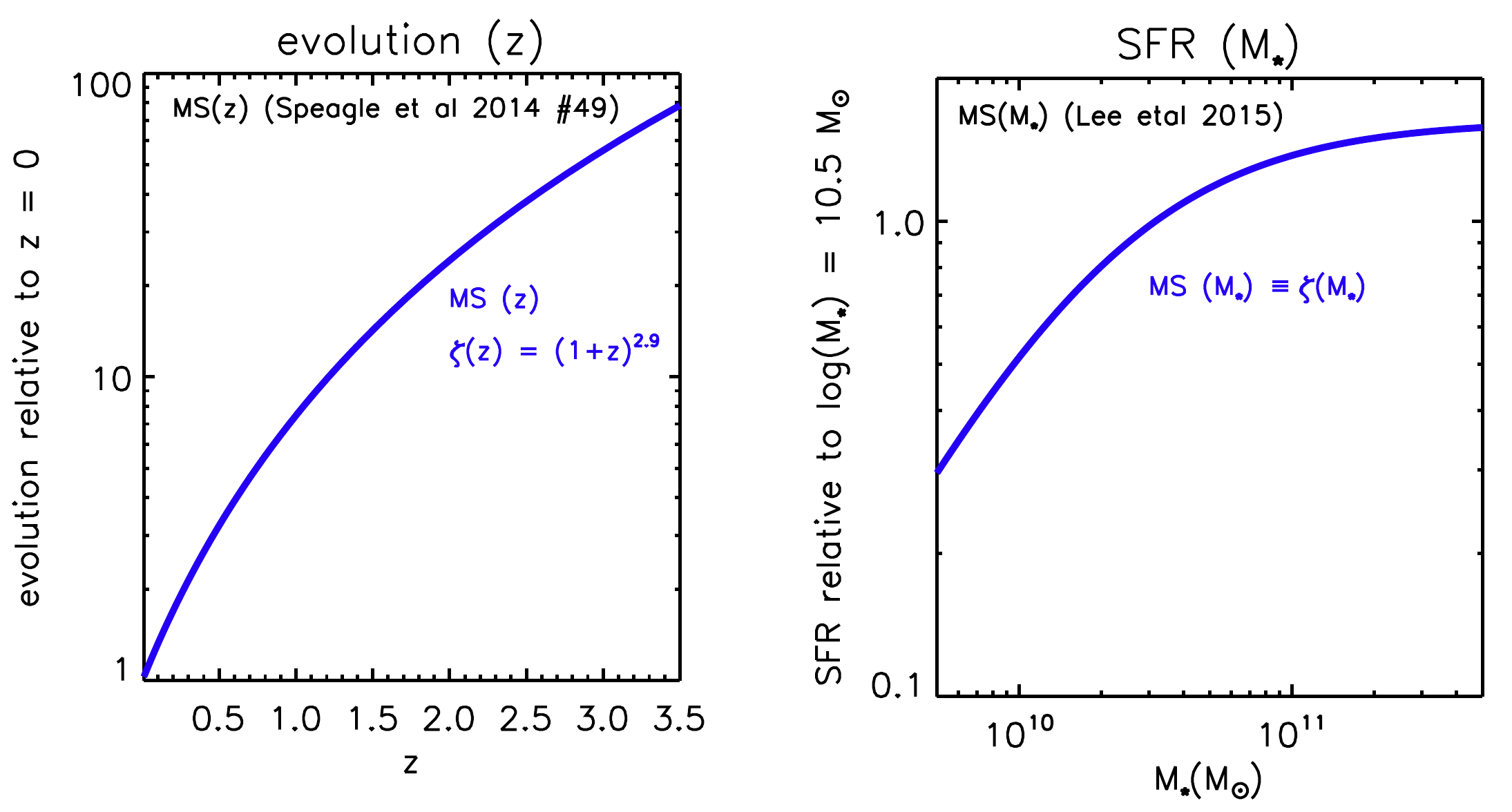}
\caption{The dependencies of the MS star formation rate as a function of redshift and stellar mass are shown on the Left and Right, respectively. The evolution of the MS as a function of z, \zfn(z) \citep[][fit \# 49]{spe14} is normalized to z = 0 --  \zfn(z) $= (1+z)^{2.9}$ for log (M$_{\rm  *})$ = 10.5 \msun. The shape of the MS as a function of mass (\zfn(M$_{\rm  *}$)) is taken from the z = 1.2 MS of \cite{lee15}, normalized to a fiducial mass of log M$_{\rm *} = 10.5$ \msun.  The constant of normalization is SFR = 3.23 \msun~$yr^{-1}$ for z = 0 and log M$_{\rm *} = 10.5$ \msun. }
\label{ms} 
\end{figure*}

\section{Evolution of the star formation rates and stellar mases}\label{basis}

The bulk of the SF galaxy population can be located on a `Main Sequence' (MS) locus in the plane of SFR versus M$_{\rm *}$ \citep{noe07,pen10,rod11,whi12,lee15}.\footnote{See \cite{eal16} for some reservations regarding the MS.} 

\subsection{Evolution of the MS with redshift}\label{ms_z}

The MS locus (SFR(z,M$_{\rm *})$) evolves to higher SFRs 
at earlier epochs as shown in Figure \ref{ms}-Left \citep[][their favored model \#49]{spe14}. This MS definition was based on an extensive  reanalysis of all previous work; similar MS definitions  have been obtained by \cite{bet12}, ~ \cite{whi14}, \cite{sch15}, \cite{lee15} and \cite{tom16}. Model \# 49 is expressed analytically by  
 \begin{eqnarray}  
  \rm ~SFR_{\rm MS} &=& ~ 10^{(0.59~\rm log(M_{\rm *})-5.77)}  \times \nonumber \\ &&(1+z)^{( 0.22~\rm log(M_{\rm *})+0.59)} ~.     \nonumber
 \end{eqnarray} 
 \noindent We use this function, evaluated at log(M$_{\rm *}) = 10.5$\msun~ to describe the redshift evolution of the MS, 
  \begin{eqnarray}  
  \rm \zfn(z) \equiv ~SFR_{\rm MS}(z) /  SFR_{\rm MS}(z=0) &=& ~ (1+z)^{2.9} ~.          \label{ms_evolution_speagle} 
 \end{eqnarray} 
 
\noindent For fitting the redshift dependence of ISM masses in Section \ref{ism_fitting}, we fit for an exponent of (1+z) so that we can directly compare the evolution of the ISM masses with that of the SFRs which have a $(1+z)^{2.9}$ dependence.

\subsection{Shape of the MS with $M_*$}\label{ms_m}

Although the early descriptions of the MS used single power laws as a function of stellar mass, the more recent work  \citep{whi14,lee15,tom16} indicated
a break in the slope of the MS at M$_{\rm *} > 5\times10^{10}$ \msun, having lower sSFR at higher M$_{\rm *}$. Here, we use the shape of the MS from \cite{lee15} at z = 1.2 
for the stellar mass dependence of the MS, 
 \begin{eqnarray}  
  \rm ~SFR_{\rm MS} &=& ~ 10^{(1.72~\rm -~ log(1 ~+~ (10^{(log M_{\rm *} -10.31)})^{-1.07}))}  \nonumber  ~     \nonumber
 \end{eqnarray} 
 \noindent and normalize to unity at  log M$_{\rm *}$ = 10.5 \msun, 
  \begin{eqnarray}  
  \rm \zfn(M_{\rm *}) &\equiv& ~\rm SFR_{\rm MS}(M_{\rm *}) /  SFR_{\rm MS}(log M_{\rm *} =10.5) ~.
\label{ms_mass} 
 \end{eqnarray} 
 
 \noindent 
 Figure \ref{ms}-Right shows the adopted shape function \zfn(M$_{\rm *}$). The normalization constant is a SFR = 3.23 \msun~$yr^{-1}$ at z = 0 and logM$_{\rm *} = 10.5$ \msun.

We have investigated the use of alternative MS formulations, and the conclusions 
derived here do not depend qualitatively on the adopted MS formulation, although the numerical values of the power law exponents can change by $\sim\pm0.1$ with adoption of one of the other MS definitions. We used the combination of \cite{spe14} and \cite{lee15}, since the former covers the complete redshift range covered here (but has only power-law
dependence on mass) whereas the latter has the break in the MS at log(M$_{\rm *}) \sim$ 10.5 \msun~ which is seen in the latest determinations of the MS.

\subsection{Continuity of the MS Evolution} \label{continuity}

In our analysis, we make use of a principle we refer to as the \emph{Continuity of Main Sequence Evolution} -- simply stated,  
the temporal evolution of the SF galaxy population may be followed by Lagrangian integration of the MS galaxy evolution. This follows from the fact that
approximately 95\% of the SF galaxies at each epoch lie on the MS with SFRs dispersed only a factor 2 above or below the MS \citep[e.g.][]{rod11}. (A similar approach has been used by 
\cite{noe07a}, \cite{ren09} and \citep{lei12} (and references therein).

This continuity assumption ignores the galaxy buildup arising from \emph{major} mergers of similar mass galaxies since they can depopulate the MS population in the mass range of interest. In fact, 
the major mergers may be responsible for some of the 5\% galaxies in the SB population above the MS (see Section \ref{sb_ms}). On the other hand, minor mergers may be considered simply as one element of the average accretion process considered in Section \ref{accretion}. 

We are also neglecting the SF quenching processes in galaxies. This occurs mainly in the highest mass galaxies (M$_{\rm *} > 2\times10^{11}$\msun ~ at z  $>1$) and in dense environments at lower redshift \citep{pen10,dar16}. At z $> 1.2$ the quenched red galaxies 
are a minor population (see Figure \ref{lilly_madau}) and the quenching processes are of lesser importance.

\begin{figure}[ht]
\epsscale{1}  
\plotone{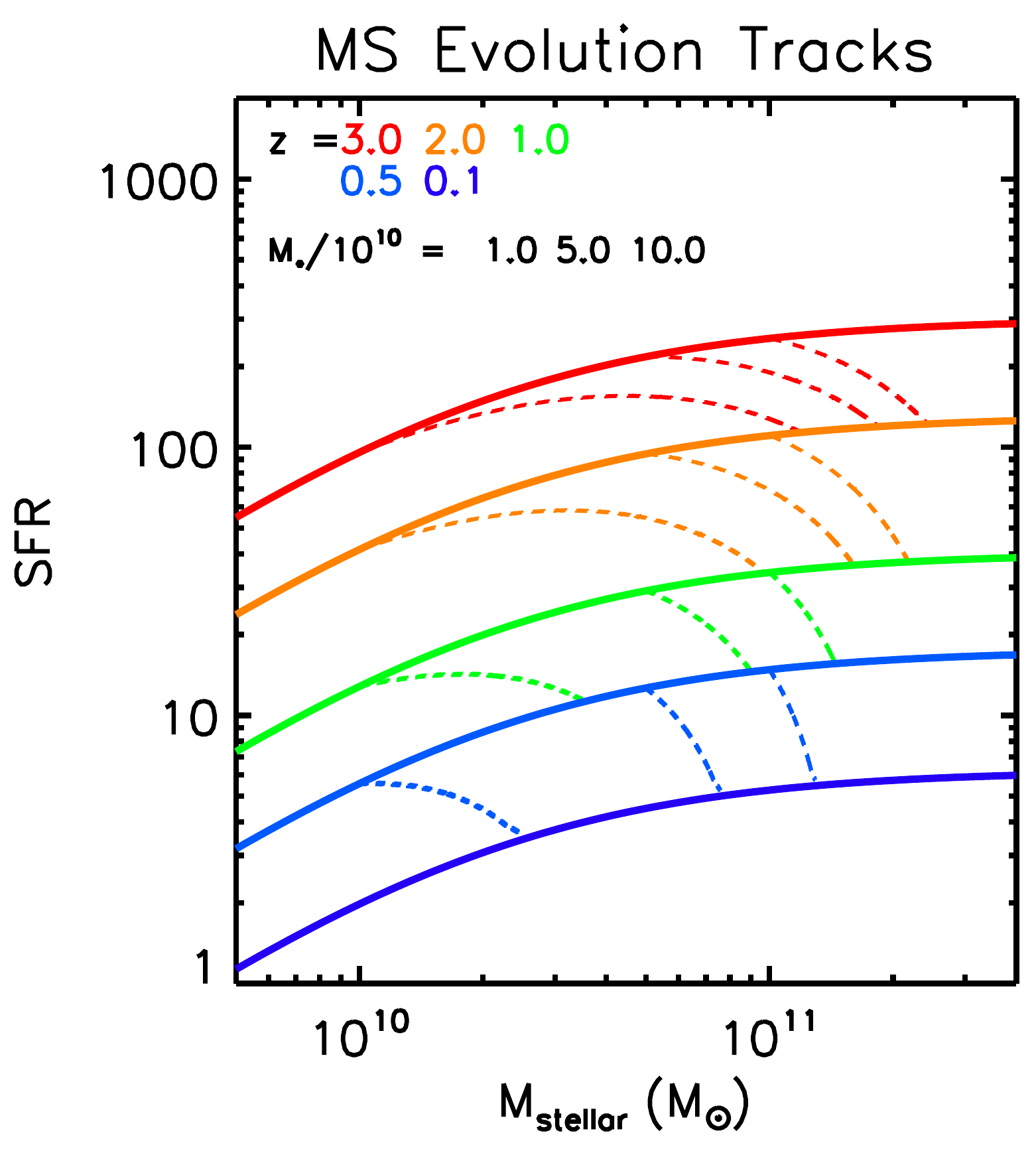}
\caption{The evolution of the star forming galaxy MS evolves to lower SFRs at lower z for all stellar masses. The curved 
downward tracks (dashed lines) show the evolution of characteristic stellar masses (1, 5, and 10 $\times10^{10}$\msun)between the MS lines at the adjacent redshifts --  assuming that each galaxy stays on the MS (z) and its increase
in stellar mass is given by integration of the SFR$_{MS}$ (M$_{\rm *}$) over time. This evolution 
is derived here using the MS fits from \cite{spe14} which are valid at z = 0.25 to 3.5 with the mass dependence for the MS taken from \cite{lee15} (see Figure \ref{ms}). To account for mass returned to the ISM from stellar mass-loss, we adopt a mass-return percentage $= 30$\% of the SFR \citep{lei11}, i.e. the stellar mass of the galaxy grows at a rate = $0.7\times$SFR.  }
\label{ms_evolution} 
\end{figure}

The paths of evolution in the 
SFR versus M$_{\rm *}$ plane can be easily derived  since the MS loci give $dM_{\rm *}/dt$ = SFR (M$_{\rm *}$). 
One simply follows each galaxy in a Lagrangian fashion 
as it builds up its mass.

Using this Continuity Principle to evolve each individual galaxy over time, the evolution for MS galaxies across the SFR-M$_{\rm *}$ plane is as shown in Figure \ref{ms_evolution}. Here we have assumed that 30\% of the SFR is eventually put back into the ISM by stellar mass-loss. This is appropriate for the mass-loss from a stellar population with a Chabrier IMF \citep[see][]{lei11}.
In this figure, the curved horizontal lines are the MS at fiducial epochs or redshifts, while the downward curves are the evolutionary tracks 
for fiducial M$_{\rm *}$ from 1 to 10 $\times10^{10}$ \msun. At higher redshift, the evolution is largely toward increasing M$_{\rm *}$ whereas at lower redshift the evolution is  
in both SFR and M$_{\rm *}$. In future epochs, the evolution is likely to be still more vertical as the galaxies exhaust their gas supplies. Thus there are three phases in the evolution:
\begin{enumerate}
\item the gas accretion-dominated and stellar mass buildup phase at z $> 2$ corresponding to cosmic age less than 3.3 Gyr (see Section \ref{accretion});
\item the transition phase where gas accretion approximately balances SF consumption and the evolution becomes  diagonal
and,
\item the epoch of ISM exhaustion at z $\lesssim$ 0.1 (age 12.5 Gyr) where the evolution will be vertically downward in the SFR versus M$_{\rm *}$ plane.
\end{enumerate} 

These evolutionary phases are all obvious (and not a new development here), but in Section \ref{accretion} we make use of the Continuity Principle to derive the accretion rates and hence substantiate the 3 phases as separated by their accretion rates relative to their SFRs. When 
these phases begin and end is a function of the galaxy stellar mass -- the transitions in the relative accretion rates take place much earlier for the more massive galaxies.

\subsection{MS Galaxy Evolution Paths}

The SFR evolution for the MS (relative to the MS at z  = 0) is shown in Figure \ref{ms}. Figure \ref{ms_evolution} shows the evolution for three fiducial stellar masses. In our analysis to fit the evolution of the ISM, SFR and accretion rates (Section \ref{fits}), we will fit directly for 1+z dependence. The derived evolutionary trends with redshift can then be
referenced to the well-known redshift evolution SFRs. 

\begin{deluxetable*}{llcccccccccc}[h]
\footnotesize
\tablecaption{\bf{ALMA Datasets used for Continuum Measurements}  }

\tablehead{ 
 \colhead{Project Code} &  \colhead{PI}  & \colhead{Band} & \colhead{$<\nu_{obs}>$} & \colhead{$\#$ pointings}  & \colhead{$<$HPBW$>$} &  \colhead{$\sigma_{rms}$} \\ 
 \colhead{} & \colhead{} & \colhead{} & \colhead{(GHz)} & \colhead{} & \colhead{(arcsec)} & \colhead{(mJy)}  }
\startdata 
\\
2011.0.00097.S & Scoville & 7 & 341 &  105 &       0.64 &      0.300 \\
2011.0.00964.S & Riechers & 7 & 296 &    1 &       0.69 &      0.091 \\
2012.1.00076.S & Scott & 6 & 244 &   20 &       1.27 &      0.062 \\
2012.1.00323.S & Popping & 6 & 226 &    2 &       1.01 &      0.093 \\
2012.1.00523.S & Capak & 6 & 291 &    3 &       0.74 &      0.041 \\
     ~~~~~~~~ ''          &  &      7 &    298 &    5 &       0.77 &      0.039 \\
2012.1.00978.S & Karim & 7 & 338 &    6 &       0.37 &      0.084 \\
2013.1.00034.S & Scoville & 6 & 245 &   60 &       0.66 &      0.063 \\
      ~~~~~~~~~''                  &  &      7 &    343 &   98 &       0.51 &      0.141 \\
2013.1.00118.S & Aravena & 6 & 240 &  128 &       1.31 &      0.124 \\
2013.1.00151.S & Schinnerer & 6 & 240 &   86 &       1.71 &      0.075 \\
2013.1.00208.S & Lilly & 7 & 343 &   15 &       1.20 &      0.129 \\
2013.1.00276.S & Martin & 6 & 290 &    1 &       1.49 &      0.029 \\
2013.1.00668.S & Weiss & 6 & 260 &    1 &       1.50 &      0.029 \\
2013.1.00815.S & Willott & 6 & 263 &    1 &       0.97 &      0.017 \\
2013.1.00884.S & Alexander & 7 & 343 &   53 &       1.29 &      0.188 \\
2013.1.01258.S & Riechers & 7 & 296 &    7 &       1.05 &      0.062 \\
2013.1.01292.S & Leiton & 7 & 344 &   45 &       1.05 &      0.253 \\
2015.1.00137.S & Scoville & 6 & 240 &   68 &       1.09 &      0.081 \\
       ~~~~~~~~ ''                  &  &      7 &    343 &  300 &       0.57 &      0.158 \\
2015.1.00695.S & Freundlich & 6 & 261 &    6 &       0.05 &      0.030 \\
  \\ 
\enddata\label{projects}
\end{deluxetable*}

\subsection{Starbursts} 

At each epoch there exists a much smaller population 
($\sim5$\% by number at z = 2) which have SFRs 2 to 100 times that of the MS at the same stellar mass. Do these starburst galaxies quickly exhaust their supply of star forming gas, thus evolving rapidly back to the MS, or are their ISM masses systematically larger so that their depletion times differ little from the MS galaxies? These SB galaxies must be either a short-duration, but common evolutionary phase for the galaxies, or of long-duration, 
but a phase not undergone by the majority of the galaxy population. Despite their small numbers, their significance in the overall 
cosmic evolution of SF is greater than 5\%, since they have 2 to 100 times higher SFRs.

\section{Datasets}\label{datasets}

\subsection{ALMA Band 6 and 7 Continuum Data}

Within the COSMOS survey field, there now exist extensive observations from ALMA for the dust continuum of high redshift 
galaxies. Here, we make use of all those data which are publicly available, in addition to our own still proprietary observations. The 18 projects are listed
in Table \ref{projects} together with summary listings of the observed bands, the number of pointings and the average frequency of observation, the synthesized 
beam sizes, and the typical rms noise. The total number of pointings in these datasets is 1011, covering a total area 0.0286 deg$^2$ or 102.9 arcmin$^2$ within the Half-Power Beam Width (HPBW).


\subsection{IR Source Catalog}

Our source finding used a positional prior: the Herschel-based catalog of far-infrared sources in the COSMOS field \citep[13597 objects from[]{lee13,lee15}. COSMOS was observed at 100 $\mu$m and 160 $\mu$m by Herschel PACS \citep{pog10} as part of the PACS Evolutionary Probe program \citep[PEP;][]{lut11}), and down to the confusion limit at 250 $\mu$m, 350 $\mu$m, and 500 $\mu$m by Herschel SPIRE \citep{gri10} as part of the Herschel Multi-tiered Extragalactic Survey  \citep[HerMES;][]{oli12}). 

In order to measure accurate flux densities of sources in the confusion-dominated SPIRE mosaics, it is necessary to extract fluxes using prior-based methods, as described in \cite{lee13}. In short, we begin with a prior catalog that contains all COSMOS sources detected in the Spitzer 24$\mu$m and VLA 1.4 GHz catalogs \citep{lef09,sch10}, which have excellent astrometry. Herschel fluxes are then measured using these positions as priors. The PACS 100 and 160 $\mu$m prior-based fluxes were provided as part of the PEP survey \citep{lut11}, while the SPIRE  250, 350, and 500 $\mu$m fluxes were measured using the XID code of \cite{ros10,ros12}, which uses a linear inversion technique of cross-identification to fit the flux density of all known sources simultaneously \citep{lee13}. From this overall catalog of infrared sources, we select reliable far-infrared bright sources by requiring at least 3$\sigma$ detections in at least 2 of the 5 Herschel bands. This should greatly limits the number of false positive sources in the catalog.

An in-depth analysis of the selection function for this particular catalog is provided in \cite{lee13}, but in short, the primary selection function is set by the 24 $\mu$m and VLA priors catalog. As with many infrared-based catalogs, there is a bias toward bright, star-forming galaxies, but the requirement of detections in multiple far-infrared bands leads to a flatter dust temperature selection function than typically seen in single-band selections.  

Since the selection function is biased to IR bright and massive galaxies, the sample is not  representative of the high redshift SF galaxy population. However, in the analytic fitting 
below we obtain analytic dependencies for the ISM masses and SFRs on the sSFR, the stellar mass and redshift. These analytic fits can then be used to analyze 
the more representative populations. This approach is used in Section \ref{cosmic} to estimate the cosmic evolution of ISM. 

\subsection{Redshifts, Stellar Masses and Star Formation Rates} 
Spectroscopic redshifts 
were used for 5066 of these sources; the remainder had photometric redshifts from \cite{ilb09} and \cite{lai15} (This catalog does 
not include objects for which the photometric redshift fitting indicated a possible AGN.) For the final sample of sources falling within the ALMA pointings, 38\% had spectroscopic redshifts.

The primary motivation for using the Herschel IR catalog for positional priors is the fact that once one has far-infrared detections of a galaxy, the SFRs can be estimated more reliably (including the dominant 
contributions of dust-obscured SF) rather than relying on optical/UV continuum estimation, which often have  corrections by factors $\gg5$ for dust obscuration.  This said, the SFRs derived from the far-infrared are still probably individually uncertain by a factor 2, given uncertainties in the stellar IMF 
and the assumed timescale over which the young stars remain dust-embedded. 

 The conversion from IR (8-1000$\mu$m) luminosity makes use of $SFR_{IR} = 8.6\times10^{-11} L_{IR}/\lsun$ using a Chabrier stellar IMF from 0.1 to 100 \msun \citep{cha03}.
The scale constant is equivalent to assuming that 100\% of the stellar luminosity is absorbed by dust for the 
first $\sim$100 Myr and 0\% for later ages. For a shorter dust enshrouded timescale of 10 Myr the scaling constant is $\sim$1.5 times larger \citep{sco13a}. In 706 of the 708 sources in our measured sample, the IR SFR was greater than the optical/UV SFR. The final SFRs are the sum of the opt/UV and the IR SFRs.  

The stellar masses of the galaxies are taken from the photometric redshift catalogs \citep{ilb09,lai15}; these are  
also uncertain by at least factors of 2 due to uncertainties in the spectral energy distribution (SED) modeling and extinction corrections. Their uncertainties are probably less than those for the optically derived SFRs, since the stellar mass in galaxies is 
typically more extended than the SF activity, and therefore is likely to be less extincted. 

The other galaxy property we wish to correlate with the derived ISM masses is 
the elevation of the galaxy above or below the SF MS. This enhanced SFR is quantified by sSFR/sSFR$_{\rm MS}(z,M_{\rm *})$ with the MS definition taken from the combination of \cite{spe14} and \cite{lee15} (see Section \ref{ms_z} and \ref{ms_m} and Figure \ref{ms}). 

\subsection{Complete Sample of ALMA-detected IR-bright Galaxies}

\emph{ The galaxy sample analyzed here includes 708 galaxies, yielding 575 to be used in the fitting after redshift and steller mass selection. This large number of objects can then overcome uncertainties in the z, SFR and M$_{\rm *}$ for each individual object.} 
This sample also has calibrations that are uniform across the full sample without the
need for zero-point corrections. All the objects are in the COSMOS survey field and thus have the same photometry and scheme for evaluating the stellar masses and the
redshifts; they also all have similar depth Herschel and Spitzer infrared observations. 

For the ISM measurements, we use exclusively continuum observations from ALMA -- these are consistently 
calibrated and with resolution ($\sim1$\arcsec) such that source confusion is not an issue. Lastly, our analysis involving the RJ dust continuum avoids the issue of variable excitation 
which causes uncertainty when using different CO transitions across galaxy samples. The excitation and brightness per unit mass for the different CO transitions is likely to vary by factors of 2 to 3 from one galaxy to another 
and within individual galaxies \citep[see][]{car13}. 

\subsection{Source Measurements}

At each IR source position falling within 
the ALMA primary beam HPBW (typically 20\arcsec~in Band 7), we searched for a significant emission source ($> 2 \sigma$) within 2\arcsec~radius of the IR source position. This radius is the expected maximum size for these galaxies. The adopted detection 
limit implies that $\sim2$\% of the detections at the $2\sigma$ limit could be spurious. Since there are $\sim240$ galaxies detected at 2-3$\sigma$, we can expect $\sim4$ of the detections could be false. \footnote{We have also used detection thresholds of 3 and 4$\sigma$
and did not find significant changes in the fitting coefficients in the analysis below, that is less than 10\% change; the 2$\sigma$ limit was therefore used since it yields measurements and uncertainty estimates for the complete 
sample of sources.} 

Some of the sources are likely to be somewhat extended relative to the synthesized beams (typically $\sim1$\arcsec); we therefore measure both the peak and integrated fluxes. The latter were 
corrected for the fraction of the synthesized beam falling outside the aperture. 
The adopted final flux for each source was the maximum of these as long as the SNR was $>2$. 

The noise for both 
the integrated and peak flux measures was estimated by placing 50 randomly positioned apertures of similar size in other areas of the FOV and measuring the 
dispersion of those measurements. The synthesized beams for most of these observations were $\sim1$\arcsec, and the interferometry should have good flux 
recovery out to sizes $\sim4$ times this; since the galaxy sizes are typically $\leq2$ to 3\arcsec, we expect the flux recovery to be nearly complete, that is, there should be relatively little resolved-out emission.  
All measured fluxes were corrected for primary beam attenuation. The maximum correction is a factor 2 when the source is near the HPBW radius for the 12m telescopes.

A total of 708 of the Herschel sources were found within the ALMA FOVs, the positions of this sample were used as priors for the ALMA flux measurements. For the sample of 708 objects, the measurements yielded 708 and 182 objects with $>2$ and $>7\sigma$ detections, respectively. Thus at 2$\sigma$, all sources were detected. No correction for Malmquist bias was applied since 
there were detections for the complete sample of sources falling within the survey area. 

We then restricted the sample used for fitting to z and M$_{*}$ where the sampling is good (see Section \ref{fits}). A total of 772 flux measurements were made -- some of the Herschel sources had multiple 
ALMA observations. In some cases the duplicate pointings were in both Bands 6 and 7 (see Appendix \ref{duals}). 

\begin{figure}[ht]
\epsscale{1}  
\plotone{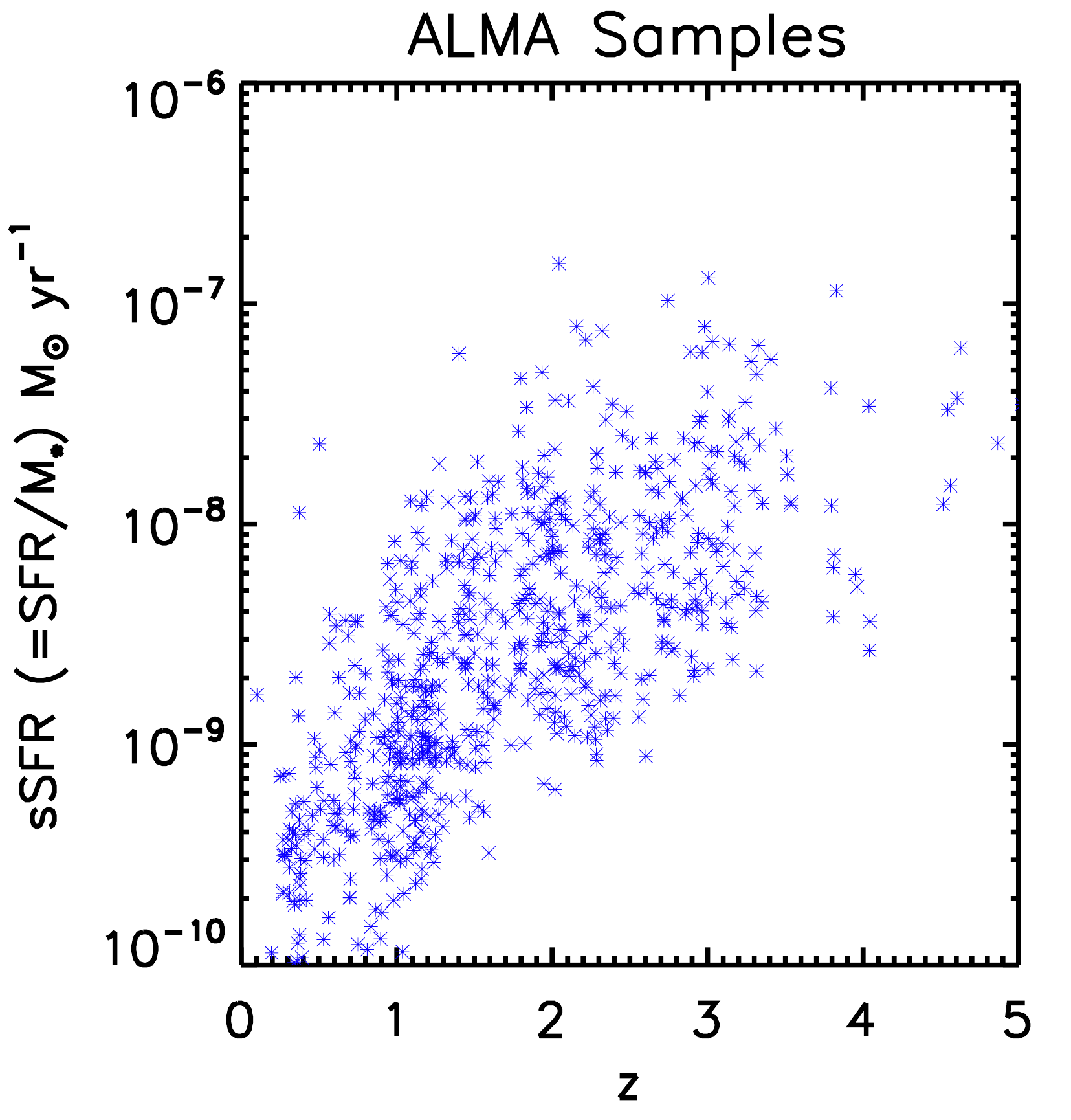}
\caption{The distributions of sSFR and redshifts for the objects which were detected are shown.}
\label{sample_ssfr} 
\end{figure}

\begin{figure}[ht]
\epsscale{1.}  
\plotone{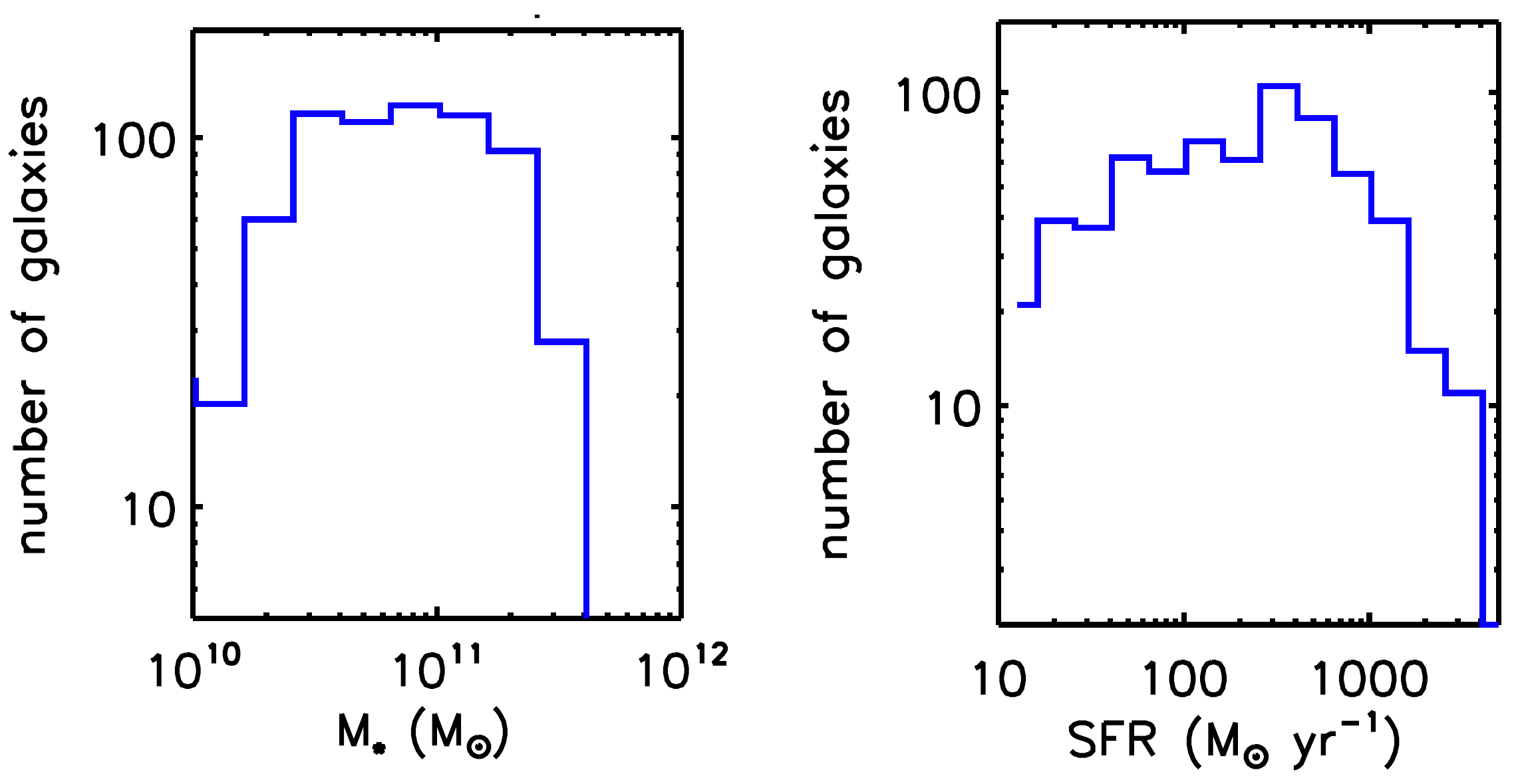}
\caption{The distributions of stellar masses and SFRs for the detected objects.}
\label{sample_m_sfr} 
\end{figure}

\begin{figure}[ht]
\epsscale{1.}  
\plotone{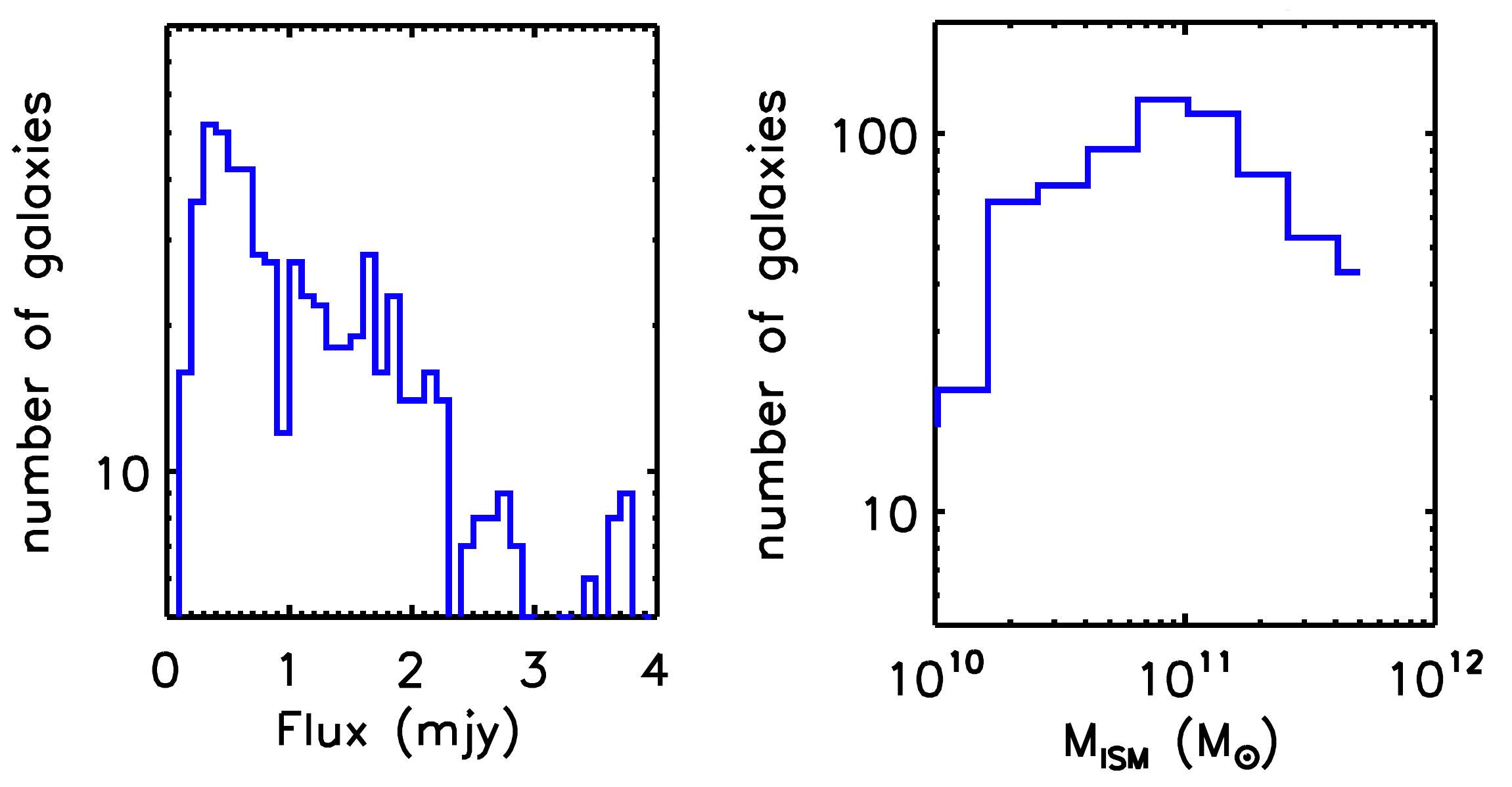}
\caption{The distributions of observed flux densities and derived ISM masses.}
\label{sample_fluxes} 
\end{figure}

In summary, all of the Herschel sources within the ALMA pointings
were detected. The final sample of detected objects with their sSFRs and redshifts is shown in Figure \ref{sample_ssfr}; the histogram distributions of M$_{\rm *}$ and SFR are in Figure \ref{sample_m_sfr}. Figure \ref{sample_fluxes} shows the distribution of measured fluxes and derived ISM masses.

 In approximately 10\% of the images there is more than one detection. However, the redshifts of these secondary sources and 
the distributions of their offsets from the primary source, indicate that most of the secondaries are not physically associated with the primary sources. 

\medskip

\medskip

\medskip

\section{ISM Masses Estimated from RJ Dust Emission}\label{rj_dust}

Here we briefly summarize the physical basis underlying the estimation of ISM masses from the Rayleigh Jeans dust emission. Appendix \ref{dust_app} provides more detail, including the full equation used for mass calculations; a complete discussion is given in \cite{sco16}.

\subsection{Physical Basis}

The far infrared dust continuum emission of galaxies arises from interstellar dust, heated by absorption of short wavelength 
radiation (UV, optical and near infrared).  The short wavelength sources of this luminosity are recently formed stars or AGN;
hence the far infrared luminosity provides a measure of that portion of the SF and nuclear activities which is dust obscured. For the standard Galactic ISM at solar metallicity, a modest column of gas and dust, $N_H \gtrsim 2\times10^{21}$ cm$^{-2}$, will produce an extinction A$_V = 1$ mag. 
Since the dust grains 
are small ($\lesssim 1$ micron), the dust opacity is 
highest in the UV and optical but decreases strongly 
at far infrared and submm wavelengths ($\kappa_{\nu}\propto \lambda^{\sim-1.8}$ at $\lambda \gtrsim 200\mu$m \citep{pla11a}). 

At long wavelengths on the Rayleigh-Jeans tail, the dust emission is almost always optically thin and the emission flux per unit mass of dust is 
only linearly dependent of the dust temperature. Thus the flux observed on the RJ tail provides a linear estimate of the dust mass 
and hence the ISM mass, provided the dust emissivity per unit mass and the dust-to-gas abundance ratio can be constrained. 
Fortunately, both of 
these prerequisites are well established from  
 observations of nearby galaxies \citep[e.g.][]{dra07b,gal11}. 
 
 Theoretical understanding of the dust emissivity has  also significantly improved in the last two decades \citep{dra07a}. On the optically thin, RJ tail of the IR emission, the observed flux density is given by 
 
\begin{eqnarray}\label{fnu}
S_{\nu} &\propto&  \kappa_{D}(\nu)  T_{\rm D} \nu^2 {M_{\rm D}\over{d_L^2}} \nonumber 
\end{eqnarray}

\noindent where $T_{\rm D}$ is the temperature of the emitting dust grains,  $\kappa_{D}(\nu)$ is the dust opacity per unit mass of dust, $M_{\rm D}$ is the total
mass of dust and $d_L$ is the source luminosity distance. Thus, the mass of dust and ISM can be estimated from observed specific luminosity $L_{\nu}$ on the RJ
tail:
\begin{eqnarray}\label{lnu}
M_{D} &\propto& { L_{\nu}  \over { <T_D>_M   \kappa_{D}(\nu)}}  \\
M_{ISM} &\propto& { L_{\nu}  \over { f_{D} <T_D>_M   \kappa_{D}(\nu)}}  .
\end{eqnarray}

\noindent Here $<T_D>_M$ is the mean mass-weighted dust temperature and $f_{D}$ is the dust-to-ISM mass ratio (typically $\sim 1/100$ for solar metallicity ISM). 

\subsection{Empirical Calibration} 

In \cite{sco16}, we used a sample of 30 local star forming spiral galaxies, 12 ultraluminous IR galaxies (ULIRGs) and 30 z $\sim 2-3$ submm galaxies (SMGs) to 
empirically calibrate the RJ luminosity-to-mass ratio 

\begin{eqnarray}
\alpha_{\nu} &\equiv& < L_{\nu_{850\mu \rm m}} /M_{\rm mol}>  \nonumber \\
&=& 6.7\pm1.7\times 10^{19} \rm erg ~sec^{-1} Hz^{-1} {\msun}^{-1} ~.
\end{eqnarray}\label{alpha}

\noindent The gas masses were in all cases estimated from global measurements of the CO (1-0) emission in the galaxies assuming a 
single Galactic CO conversion factor $\alpha_{CO} = 6.5 ~\msun / \rm{K~km~s^{-1} pc^2}$ or $X_{CO} = 3\times10^{20}$ N(H$_2$) cm$^{-2}$ (K km s$^{-1})^{-1}$. 
These molecular gas masses include a factor 1.36 to account for the associated mass of heavy elements (mostly He at 8\% by number).

 It is noteworthy that across the three diverse samples of galaxies the empirical calibration varied less than a factor 2 (see Figure \ref{empir_cal}). The calibration 
therefore includes only massive galaxies (Milky Way or greater) which are likely to have near solar metallicity.
 The samples of high redshift galaxies in the present paper are therefore restricted to stellar mass $M_* > 2\times10^{10}$ \msun (and most are above $5\times10^{10}$ \msun). 
 
 Probing lower mass galaxies, which presumably would have significantly sub-solar metallicity, will require careful calibration as a function of metallicity or mass. We note that in \cite{dra07b}, there is little evidence of variation in the dust-to-gas abundance ratio for the first factor of 4-5 down from solar metallicity. However, at lower metallicities the 
dust-to-gas abundance does clearly decrease (see Figure 17 in \cite{dra07b} and Figure 16 in \cite{ber16}). 

Lastly, we point out that since the standard Galactic ISM dust-to-gas abundance ratio (quantified by $2\times10^{21} \rm H ~cm^{-2} / \rm A_V$) is generally the same 
in both molecular and atomic hydrogen regions; one might expect that a similar calibration $\alpha_{\nu}$ pertains to the atomic and molecular phases of galaxies, 
provided one restricts to approximately solar metallicity HI and H$_2$ regions in the galaxies, such as the inner disks. For this reason, we will henceforth refer to our mass estimates as M$_{ISM}$. The 
ISM mass estimates found below for the high z galaxies are very large (5 - 100$\times$ that of the Galaxy); one could  
expect that the dominant phase is in fact molecular within the high-z galaxies. 

\subsection{Mass-weighted $\rm T_D$} \label{rj_td_dust}

It is important to emphasize that the dust temperature entering Equation \ref{lnu} is a mass-weighted dust temperature $\rm <T_D>_M$, not the 
luminosity-weighted temperature $\rm <T_D>_L$. The latter is what would be derived by fitting the overall SED of the emergent IR radiation to 
a black body or modified black body curve. In such spectral fitting, the derived T$_D$ is largely determined by the shape and location of the 
SED peak, and hence the dust emitting the bulk of the \emph{emergent} luminosity. 

Often, the far infrared SEDs are analyzed by fitting either modified black body curves or libraries of dust SEDs to the observed SEDs \cite[e.g.][]{dra07b,dac11,mag12,mag12a}. 
In essentially all instances the intrinsic SEDs used for fitting are taken to be optically thin. They thus do not include the attenuation expected on the short wavelength side of the far infrared 
peak, instead attributing the drop at short wavelengths solely to a lack of high temperature grains. The $\rm T_D$ determined in these cases is thus not even a luminosity-weighted $\rm T_D$ of all the dust, but just the dust above $\tau_{\nu} \sim 1$. 

Even if the source were optically thin in the mid infrared, it should be clear that the temperature determined by SED fitting will be the luminosity-weighted temperature of grains 
undergoing strong radiative heating, i.e. those closest to the sources of luminosity -- not a linear sampling of the mass of ISM in each cloud, the bulk of which is 
 at lower temperature. To illustrate the difference between mass and luminosity weighted T's, we refer to resolved studies of local GMCs: Orion \citep{lom14},
 Auriga \citep{har13} and W3 \citep{riv13}. 
 
 \subsubsection{Local GMCs and Galaxies}
 
For the two Orion GMCs, the dominant infrared luminosity sources are the KL nebula/Orion A and Orion B. \cite{lom14} present a comprehensive analysis of Herschel and Planck data. The peak 
dust temperatures are $\rm T_D \sim 40-50$ K at IRAS and Herschel angular resolutions.  
Figures 2 and 7 in \cite{lom14} clearly show that these areas of peak dust temperature are confined to the immediate vicinity of the two OB star formation regions ($\sim 2$ arcmin in size) while
the bulk of the GMCs, extending over 6 and 4\deg at $\lambda > 200\mu$m,  
is at $\rm T_D = 14 - 20$ K. A similar dichotomy between the luminosity- and mass-weighted T$_D$ is seen clearly in the Herschel data for the W3 GMC complexes \citep{riv13}.  

\cite{har13} provide a quantitative analysis of the relative masses of low and high T$_D$ dust based on their Herschel observations of the Auriga-California GMC . In this GMC, the major luminosity source 
is LkH$\alpha$101, which is the dominant source at 70$\mu$m, has $\rm T_D$ = 28K, and extends over $\sim10$\arcmin. On the other hand the $\lambda = 250\mu$m (SPIRE) image shows the GMC extending $\sim 6$\deg with a median dust temperature $\rm T_D =14.5$ K \citep{har13}. Only $\sim2$\% of the GMC mass is contained in the higher temperature region associated with LkH$\alpha$101.  

Global SED fits for 11 Kingfish Survey galaxies using two temperature components were done by \cite{gal11}. 
The warm components were 51 - 59 K and the cold components had fitted T = 17 - 24 K and the fraction of the 
total far infrared luminosity in the warm component was 21 - 81\%, with a mean value $\sim50$\% \citep[see also ][]{ski11,dal12}. 

Unfortunately, the extragalactic observations do not have the spatial resolution to resolve hot and cold dust regions in GMCs 
in other galaxies, and thus determine the relative {\emph masses} of hot and cold ISM dust. One is thus forced to learn from the 
lessons of the above cited studies of Galactic GMCs or apply physical understanding. 

With regard to the latter, the reason for the 
localization of the hot dust emitting most of the far infrared luminosity but containing little of the mass, is due to the strong wavelength 
dependence of the dust opacity. The short wavelength primary photons from the energy sources are absorbed in the first column of A$_V \simeq$ 1 mag or $2\times10^{21}$ H cm$^{-2}$. On the other hand, the luminosity of the re-radiated secondary or tertiary photons 
will require much larger columns of gas and dust to be absorbed, due to their longer wavelength  
(since the dust absorption coefficient is much lower at the longer wavelength). Thus, it becomes obvious that the mass of the cold dust component will be 
10-100 times larger than that of the warm dust. This must be the case unless the primary luminosity sources are uniformly 
distributed on the scale of A$_V \sim 1$ within the clouds  -- which is clearly not the case in Galactic GMCs. One can't rule out the possibility that the luminosity sources in the high z galaxies are more uniformly distributed 
on scales of A$_V \sim 1$ mag; however, it would seem implausible that all the SF in high z galaxies occurs in such small columns of gas.

 \subsubsection{Adopted constant $\rm T_D = 25 $K}

Here we have adopted a constant value for the mass-weighted temperature of the ISM dust to be used in our estimations 
of ISM masses. This value is slightly higher than the mean of 18 K determined by Planck observation of the cold dust 
in the Galaxy \citep{pla11a,pla11b}; we use the higher value in recognition of the fact that the mean radiation fields are likely
to be somewhat higher in the high z galaxies. Our adoption of a constant value is different from the approach used 
by others to estimate ISM dust and gas masses \citep{dac11,mag12,mag12a,gen15,sch16,ber16}. 

For the reasons 
outlined just above, we believe that using a $\rm T_D$ derived from the SED fits yields a luminosity-weighted T$_D$, not the 
physically correct mass-weighted $\rm T_D$. That is not to say that there will not be some variations in the mass-weighted 
$\rm T_D$; however, with extragalactic observations which do not resolve the star forming regions, it is impossible to determine 
the mass-weighted T$_D$. It is expected that variations in the mass-weighted $\rm T_D$ will be much less than the variations in the luminosity-weighted 
$\rm T_D$. The latter is dependent on the concentration of energy sources, which may vary considerably, whereas an increase in the overall mass-weighted 
$\rm T_D$ requires both a higher concentration of primary energy sources, \emph{and} that they be uniformly distributed on scales comparable with the 
mean free path of the primary dust-heating photons. 

Lastly, it is important to realize that the linear dependence on $\rm T_D$ for the flux to mass conversion and the expected small 
dynamic range (at most 15 - 40 K) in the mass-weighted $\rm T_D$ implies that the conversion factor will not have large 
errors with the adopted constant $\rm T_D = 25$ K.

\section{Observational Results}\label{results}
\begin{figure}[ht]
\epsscale{1.}  
\plotone{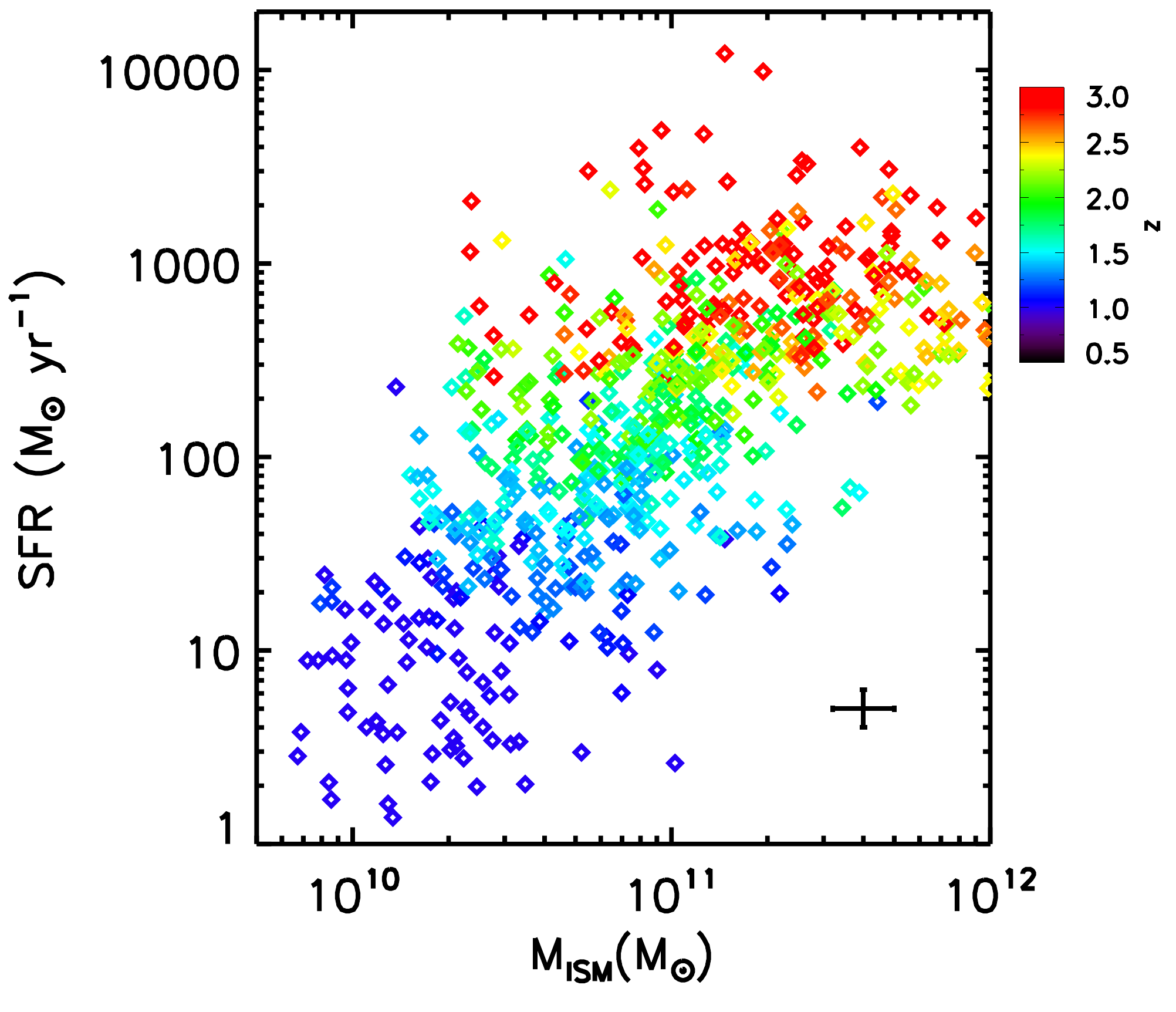}
\caption{The SFRs and derived $M_{\rm ISM}$ are shown for 708 galaxies detected in the ALMA observations. Uncertainties in both quantities 
range from 10\% to 50\%; we show a typical error bar of 25\% in the lower right corner. The observed spread in both x and y is much larger than this uncertainty, indicating 
that there must be other dependencies than a simple one-to-one correspondence between SFRs and ISM masses. }
\label{sfr_ism} 
\end{figure}

The measured ALMA fluxes were converted to estimates of the ISM masses using Equation \ref{dust_eq}. The distributions of flux and derived M$_{ISM}$ are shown in Figure \ref{sample_fluxes}.
The estimated M$_{\rm ISM}$ are shown along with the SFRs in Figure \ref{sfr_ism} for the 708 detected sources. Here one clearly sees
a strong increase in SFRs for galaxies with large M$_{\rm ISM}$; however, there is also a very large dispersion, \emph {implying 
that the SFR per unit ISM mass depends on other variables}. For example, as clearly seen in the color coding (redshift) of the 
points, there must be strong dependence on z. Yet, this can not be the entire explanation since the high redshift galaxies 
with a large range of ISM masses exhibit similarly high SFRs (that is, a lot of the galaxies in each redshift range are distributed horizontally
over a range of M$_{\rm ISM}$).  For a given SFR there are big spreads in sSFR relative to the MS and in stellar mass. The other variables linking SF and gas contents are: the stellar masses and the elevation above the MS. 
We fit for all of these dependencies simultaneously in the following sections.

\begin{figure*}[ht]
\epsscale{1.}  
\plotone{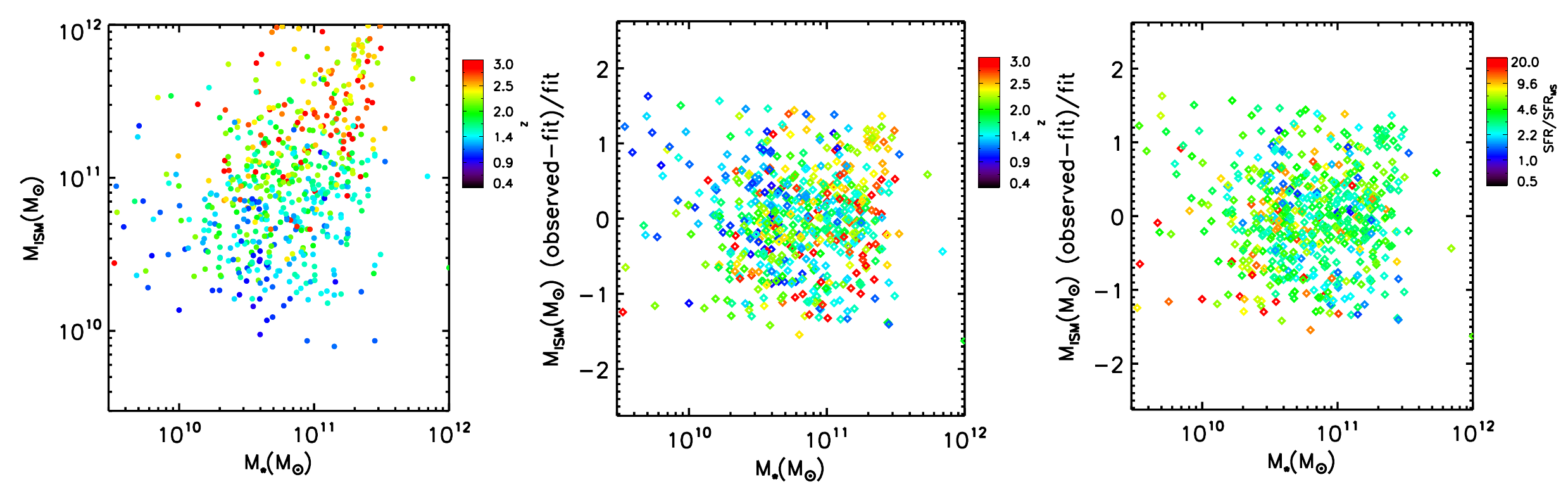}
\smallskip
\caption{The observed ISM masses are shown in the Left panel. The fractional 
differences in the fit (Equation \ref{ism_fit}) are shown in the Middle and Right panels with color coding by z and sSFR/sSFR$_{MS}$, respectively. The latter two
panels allow one to see that there are no systematic offsets with respect to either of these parameters.}
\label{fit1} 
\end{figure*}

\section{Dependence of M$_{\rm ISM}$ and SFR on z, M$_{\rm *}$, and sSFR/sSFR$_{MS}(\rm M_{\rm *},z)$}\label{fits}

In order to unravel the dependencies of ISM content and SFR on intrinsic and extrinsic galaxy parameters, we have
fit a power-law dependence of M$_{\rm ISM}$ on the most likely parameters: z, M$_{\rm *}$, and MS-Ratio ($\rm =sSFR/sSFR_{\rm MS}$); then we
use this power law expression for M$_{\rm ISM}$ in order to elucidate the dependences of the SFRs/M$_{\rm ISM}$ on z, $M_{\rm *}$, and the MS-Ratio. 
In this second stage of fitting, the terms can be viewed as SFR efficiencies per unit mass of ISM gas, as they depend separately on z, $M_{\rm *}$, and the MS-Ratio.

To avoid areas of z and stellar mass where the sampling is low, we restrict our sample further to z = 0.3 to 3 and M$_{\rm *} > 3\times10^{10}$ \msun, thus reducing the areas where Malmquist bias could be significant. The upper redshift 
limit is to avoid galaxies for which the flux measurement would be off the RJ tail; the lower mass limit is to avoid galaxies with expected low metallicity. The final 
sample used for fitting the dependencies has 575 distinct galaxies.

In all of the empirical fittings, we adopt a power law in each of the independent variables:
  \begin{eqnarray}  
\rm M_{\rm ISM}~and~ SFR &=& \rm constant~ P1^{\alpha} \times P2^{\beta} \times P3^{\gamma} \nonumber
 \end{eqnarray}  
\noindent and solve for the minimum chi-square fit as a function of the independent variables (P). A Monte Carlo Markov Chain (MCMC) routine (MLINMIX\_ERR) (in IDL) was used for the fitting. This is a Bayesian method for linear regression 
  that takes into account measurement errors in all variables, as well as their intrinsic scatter. The Markov chain Monte Carlo algorithm (with Gibbs sampling) is 
  used to randomly sample the posterior distribution \citep{kel07}.

\subsection{M$_{\rm ISM}$}\label{ism_fitting}

\begin{figure*}[ht]
\epsscale{1.}  
\plotone{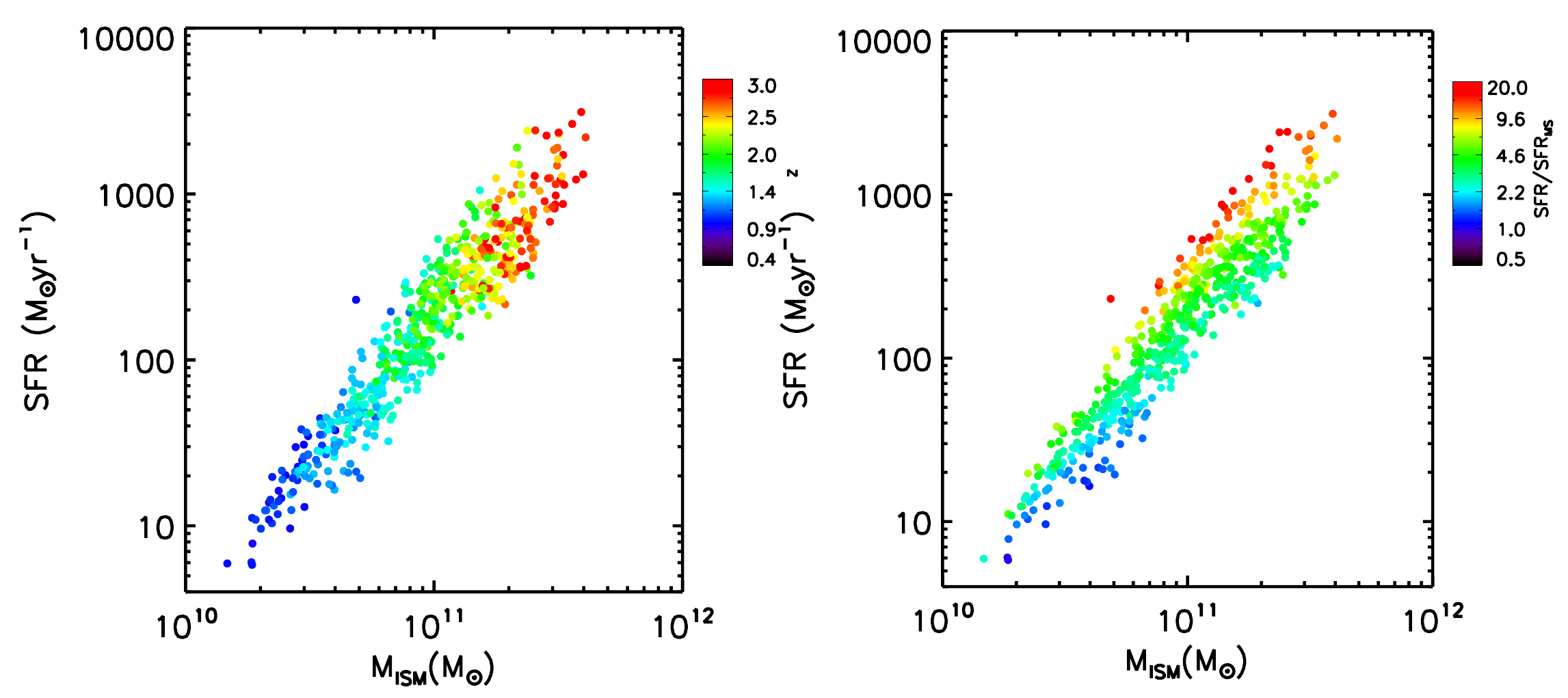}
\caption{The fitted ISM masses and observed SFRs  are shown with color coding based on redshift and sSFR relative to the MS. The dominant 
dependence of the SFR is on M$_{\rm ISM}$, with relatively weaker dependencies on z and the MS-Ratio.}
\label{sfr_ism_fig} 
\end{figure*}

The result of the MCMC fitting for the dependence of the M$_{\rm ISM}$ on redshift, MS-Ratio and stellar mass is:

 \begin{eqnarray}  
\rm M_{\rm ISM} &=& \rm 7.07\pm0.88~\times 10^9~\msun  \nonumber \\ %
&& \times (1+z)^{~1.84\pm0.14}   \nonumber \\ 
&& \times (\rm sSFR/sSFR_{MS})^{0.32\pm0.06}  \nonumber \\ 
&& \times \left( {\rm M_{\rm *}\over{10^{10} \msun}}\right) ^{0.30\pm0.04}. \label{ism_fit} 
 \end{eqnarray} 

\noindent   The uncertainties were derived from the MCMC fitting, they do not account for possible calibration uncertainties such as the conversion from $L_{IR}$ to SFR. 
In Appendix \ref{mcmc}, we provide plots showing the covariance of the fitted parameters with their distributions (see Figure \ref{cov}-Left). The posterior distributions 
of the different parameters in the MCMC fitting are single-valued (i.e., non-degenerate) and smooth. This also 
indicates that there are no degenerate parameters and hence that the choice of variables is appropriate.

In Figure \ref{fit1} the observed ISM masses (filled dots) are shown in the Left panel. 
Their fractional differences between the fit and the observations are shown in Middle and Right panels. Figure \ref{fit1}-Middle and Right show a scatter roughly equivalent to the observed values; however, given 
the large dynamic range (a factor $\sim100$) in the M$_{\rm ISM}$ and the fitting parameters, this scatter is within the combined uncertainties of those parameters. 

Thus Equation \ref{ism_fit} quantifies three major results regarding the ISM contents and their variation for high redshift galaxies relative to low redshift galaxies:
\begin{enumerate}
\item The ISM masses clearly increase toward higher z, depending on  $(1+z)^{1.84} = ((1+z)^{2.9})^{0.63}$, that is, not evolving as rapidly as the SFRs  which 
vary as $(1+z)^{2.9}$.
\item Above the MS, the ISM content increases, but not as rapidly as the SFRs (0.32 versus unity power laws).
\item The ISM contents increase as $M_{\rm *}^{0.30}$, indicating that the gas mass fractions must decrease in the higher $M_{\rm *}$ galaxies. 
\end{enumerate}

The first conclusion clearly implies the SF efficiency per unit gas mass must increase at high redshift (as discussed below). The second conclusion indicates that the galaxies above the MS have higher gas contents, but not in proportion to their elevated SFRs; and the third conclusion indicates that higher stellar mass galaxies are relatively gas-poor.
Thus, galaxies with higher stellar mass likely use up their fuel at earlier epochs and have lower specific accretion rates (see Section \ref{accretion}) than the low mass galaxies. 
This is a new aspect of the `downsizing' in the cosmic evolution of galaxies.

\subsection{SFR}

In fitting for the SFR dependencies, we wish to clearly distinguish between the 
obvious intuition that when there is more ISM there will be both more SF and a higher \emph{efficiency} for converting the 
gas to stars. Thus, we impose a linear dependence of the SFR on M$_{\rm ISM}$, using M$_{\rm ISM}$ taken from Equation \ref{ism_fit}
 rather than going back to the observed M$_{\rm ISM}$ values. Effectively, we are then fitting for the {\emph star formation efficiencies} ($\rm SFR/M_{ISM}$) for star formation per unit gas mass as a function of 
z, MS-Ratio and M$_{\rm *}$. The use of ISM masses from Equation \ref{ism_fit} is necessary in order to isolate the efficiency variation with redshift, MS-Ratio 
and M$_{\rm *}$ from the variation of the ISM masses with the same three parameters. The result is shown in Figure \ref{sfr_ism_fig} with color codings by redshift and MS-Ratio. 
The plots indicate that the dispersion in the fit is distributed around unity and there appear to be no systematic offsets with respect to redshift or MS-Ratio. 

\medskip
\begin{figure}[ht]
\epsscale{1}  
\plotone{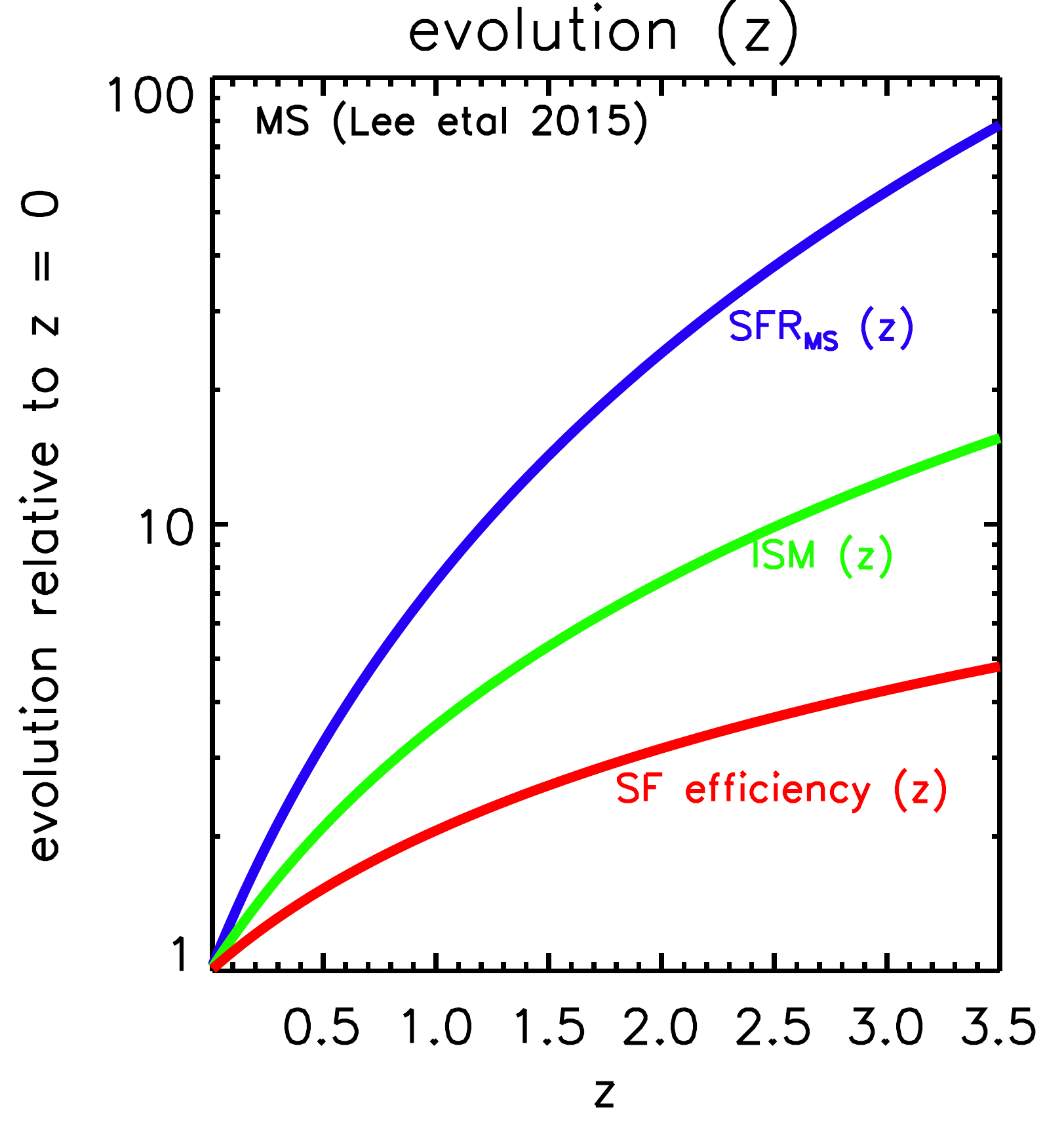}
\caption{The evolutionary dependence of the SFRs (blue), the ISM masses (green) and the SF efficiency (red) per unit mass of ISM
on the MS at a characteristic stellar mass of $5\times10^{10}$ \msun.}
\label{evolutionary_dependence} 
\end{figure}

The MCMC solution is:

 \begin{eqnarray}  
\rm SFR  &=& \rm 0.31\pm0.01~\msun~yr^{-1} \times  \left({M_{\rm ISM}\over{10^9 \msun}}\right)  \nonumber \\
&& \times   \left(1+z  \right) ^{1.05\pm0.05}  \nonumber \\ 
&&   \times \left( \rm sSFR/sSFR_{MS}\right) ^{0.70\pm0.02}  \nonumber \\ 
&&  \times \left( {\rm M_{\rm *}\over{10^{10} \msun}}\right) ^{0.01\pm0.01}. \label{sfr_fit} 
 \end{eqnarray} 

\noindent  The uncertainties in each of the fit parameters were derived from the MCMC fitting (see Appendix \ref{mcmc}). The covariances for the SFR fit are shown 
in Figure \ref{cov}-Right. The posterior distributions 
of the different parameters in the MCMC fitting are well-behaved.

Figure \ref{sfr_ism_fig} shows the fitted ISM masses (Equation \ref{ism_fit}) and measured SFRs. In the Left panel the color coding is according to the redshift, and in the Right according to their sSFR relative to the MS. Here one can see that the dominant dependence is on the ISM masses, but there is clearly dependence on redshift and sSFR.

Equation \ref{sfr_fit} was written in a form such that the three power law terms can be interpreted as efficiencies (SFR per unit gas mass). The resulting solution (Equation \ref{sfr_fit}) indicates that:
\begin{enumerate}
\item The SFR per unit ISM mass must increase approximately as $\rm (1+z)^1$ as compared to the $\rm (1+z)^{2.9}$ dependence of the SFRs, i.e. the efficiency for converting 
gas to stars (the SFR per unit mass of gas) is clearly increasing at high-z. [The increase in ISM contents is separately represented by the linear term $\rm M_{\rm ISM}$.] 
\item The SFR efficiency similarly increases as the 0.7 power of sSFR/sSFR$_{MS}$, implying that the enhanced SFRs above the MS are partially due to higher SF 
efficiencies (SFR per unit gas mass) in those galaxies -- equivalently, the SB galaxies have a shorter gas depletion 
timescale.
\item There is no significant dependence of the SFR efficiency on stellar mass ($M_*^{0.01}$). 
\end{enumerate}

The lack of dependence of the SF efficiency on galaxy mass is reasonable. If at high redshift the 
SF gas is in \emph{self-gravitating} GMCs (as at low z); the internal structure of the GMCs then influences the physics of the SF and the gas does not know 
that it is in a more or a less massive 
galaxy. 

Figure \ref{evolutionary_dependence} shows the relative evolutionary dependencies of the SFRs, the ISM masses and the SF efficiency per unit 
mass of ISM gas normalized to unity at z = 0.
The fundamental conclusion is that \emph{the elevated rates of SF activity at both 
high redshift and above the MS are due to both increased gas contents and increased efficiencies 
for converting the gas to stars.} 

\begin{figure*}[ht]
\epsscale{1.}  
\plottwo{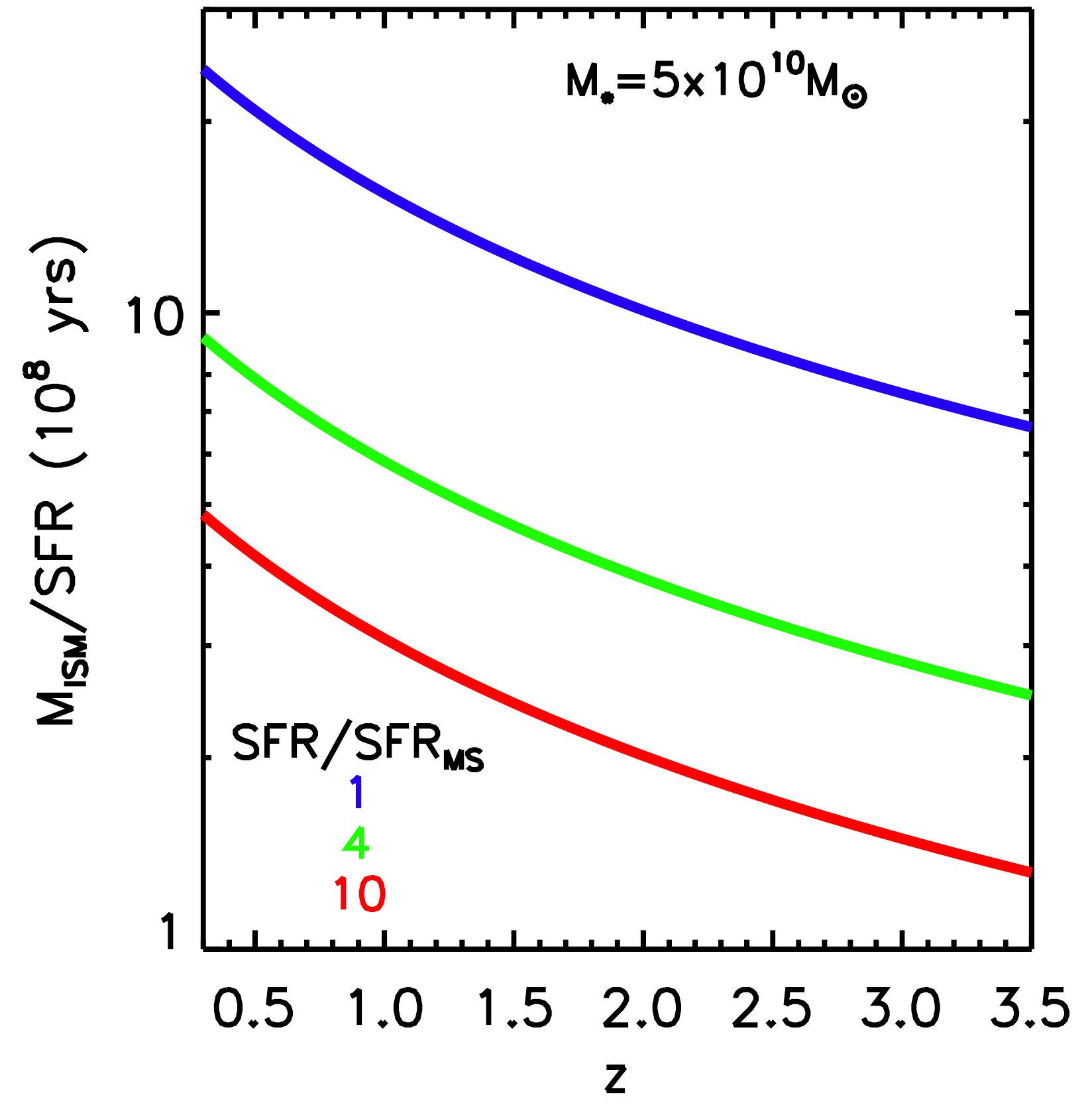}{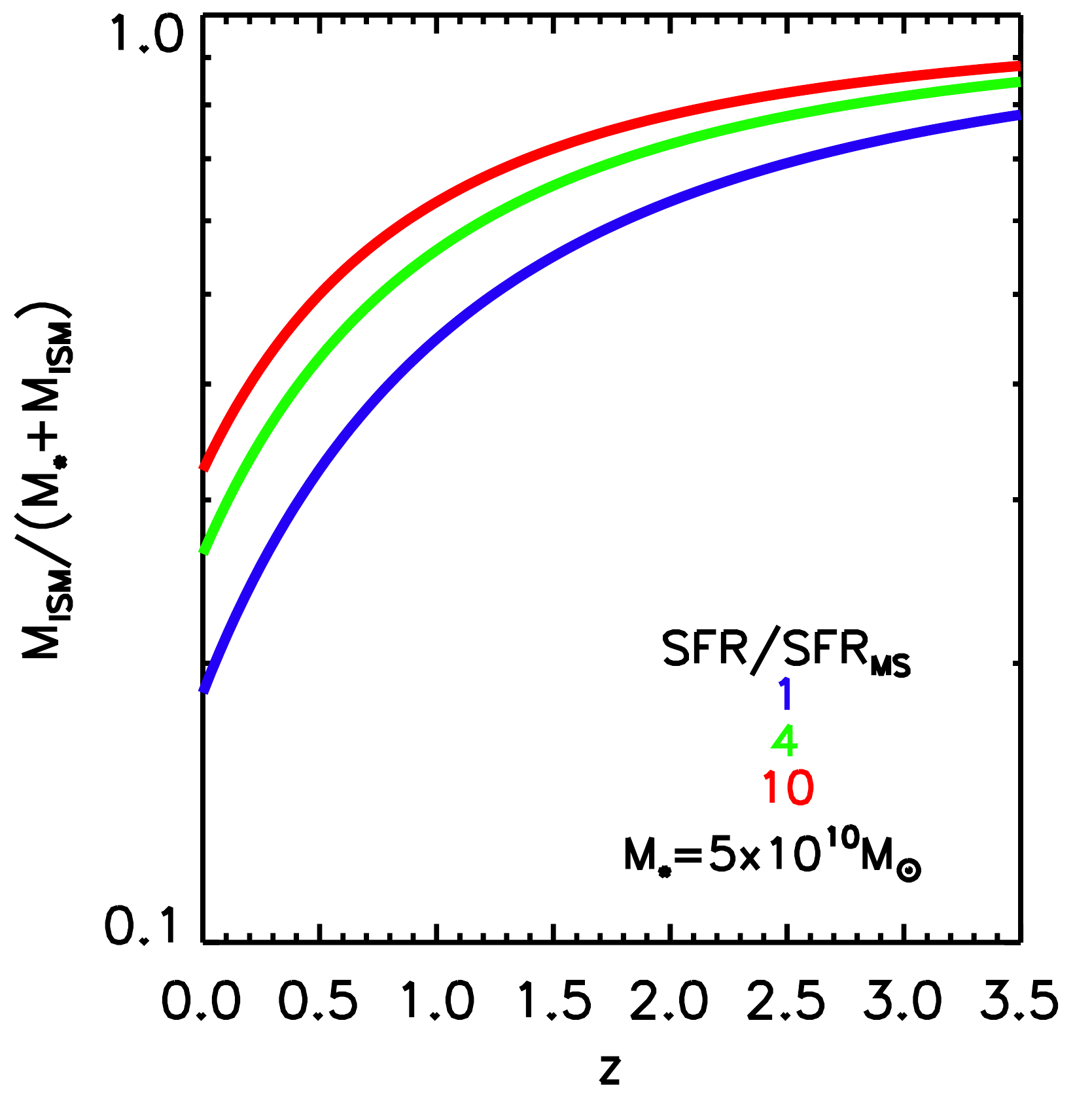}
\caption{Left - The gas depletion times (M$_{\rm ISM}$/SFR) obtained from combining Equations \ref{ism_fit} and \ref{sfr_fit}. 
Right - The gas mass fraction obtained from Equation \ref{ism_fit}. Both are shown for a fiducial mass M$_{\rm *} = 5\times10^{10}$ \msun~
and with sSFR = 1 (blue), 4 (green) and 10 (red) times that of the MS. The extrapolation of the ISM mass fraction to z = 0 is probably higher by a factor $\sim$2, compared to published values.}
\label{dep} 
\end{figure*}

\section{Implications of the ISM and SFR Relations}\label{relations}

Using the empirically based relations for M$_{\rm ISM}$ and the SFR per unit ISM mass (Equations \ref{ism_fit} and \ref{sfr_fit}), we now explore the consequences 
of these relations as obtained by simple algebraic manipulation. 

\subsection{Gas Depletion Timescales and Gas Mass Fraction}

Combining Equations \ref{ism_fit} and \ref{sfr_fit}, one can derive the gas depletion time ($\tau_{\rm dep} = \rm M_{\rm ISM}$ / SFR) and the gas mass fraction ($f_{gas}$= M$_{\rm ISM}$/(M$_{\rm *}$ + M$_{\rm ISM}$):

 \begin{eqnarray} 
 \rm  \tau_{dep}  &\equiv&  M_{ISM} / SFR \nonumber \\
 &=& \rm 3.23\pm0.10 ~ Gyr~  \times  \rm (1+z)^{-1.05\pm0.05} \times \nonumber \\ && (\rm sSFR/sSFR_{MS})^{-0.70\pm0.02} \times  M_{10}^{-0.01\pm0.01} ~,  \label{dep_fit} 
   \end{eqnarray}
 \begin{eqnarray} 
 \rm  M_{ISM}/M_{\rm *}  &=& \rm 0.71\pm0.09 ~  \times \rm (1+z)^{1.84\pm0.14} \times  \nonumber \\ && (\rm sSFR/sSFR_{MS})^{0.32\pm0.06} \times  M_{10}^{-0.70\pm0.04} ~  \label{gas_m_fit}  
    \end{eqnarray}
    \noindent and 
     \begin{eqnarray} 
\rm  f_{gas} &\equiv& \rm {M_{\rm ISM} \over{M_{\rm *} + M_{\rm ISM}}} \nonumber \\ &=&  \{ 1 ~+~1.41\pm0.18\times \rm (1+z)^{-1.84\pm0.14} \times \rm   \nonumber  \\ && (\rm sSFR/sSFR_{MS})^{-0.32\pm0.06} \times  M_{10}^{0.70\pm0.04} \}^{-1}  \label{fgas_fit} 
  \end{eqnarray}

\noindent where $\rm M_{10} = M_{\rm *} / 10^{10} \msun$. Note that Equation \ref{dep_fit} is obtained simply by canceling out the leading M$_{\rm ISM}$ term in Equation \ref{sfr_fit} (since the latter already has a linear term in $\rm M_{\rm ISM}$) -- not by dividing the full Equation \ref{ism_fit} by \ref{sfr_fit}. 

$ \tau_{dep}$ and $f_{gas}$ are shown in Figure \ref{dep}. (In keeping with convention, these depletion times do not include a correction for the mass return to the ISM during stellar evolution.) 
In Table \ref{equations2}, we have normalized these equations to z = 2 and M$_{\rm *} = 5\times10^{10}$ \msun (instead of z = 0 used earlier).
 
 Depletion times at z $> 1$ are $\sim(2 - 10)\times10^8$ yrs, much shorter than for low-z galaxies (e.g. $\sim2\times10^9$ yrs for the Milky Way). The depletion times have no dependence on $M_{\rm *}$.

The gas mass fractions shown in Figure \ref{dep}-Right are very high ($>50$\%) at z $> 2$ for SF galaxies, systematically decreasing at later epochs. They
also show a strong dependence on stellar mass ($\propto M_*^{-0.7}$, Equation \ref{gas_m_fit}) and whether a galaxy is on or above the MS (see Figure \ref{dep}-Right). It is therefore clear that there is no single, gas mass-fraction varying solely as a function of redshift (that is,  independent of the sSFR relative the MS and M$_*$). 

For MS galaxies \cite{sai13} and \cite{gen15} obtain slightly lower depletion times $\sim450$ Myr and gas mass fractions $\sim40$\% at z $\sim 2.8$. Above the MS, they find  increased ISM masses and SF efficiencies in agreement with the results here. Their results are consistent with the estimates shown in Figure \ref{dep}. 
But, it is not really appropriate to quote single value estimates for the gas mass fraction at each redshift, neglecting the sSFR and M$_*$ dependencies.

\begin{figure*}[ht]
\epsscale{1.}  
\plotone{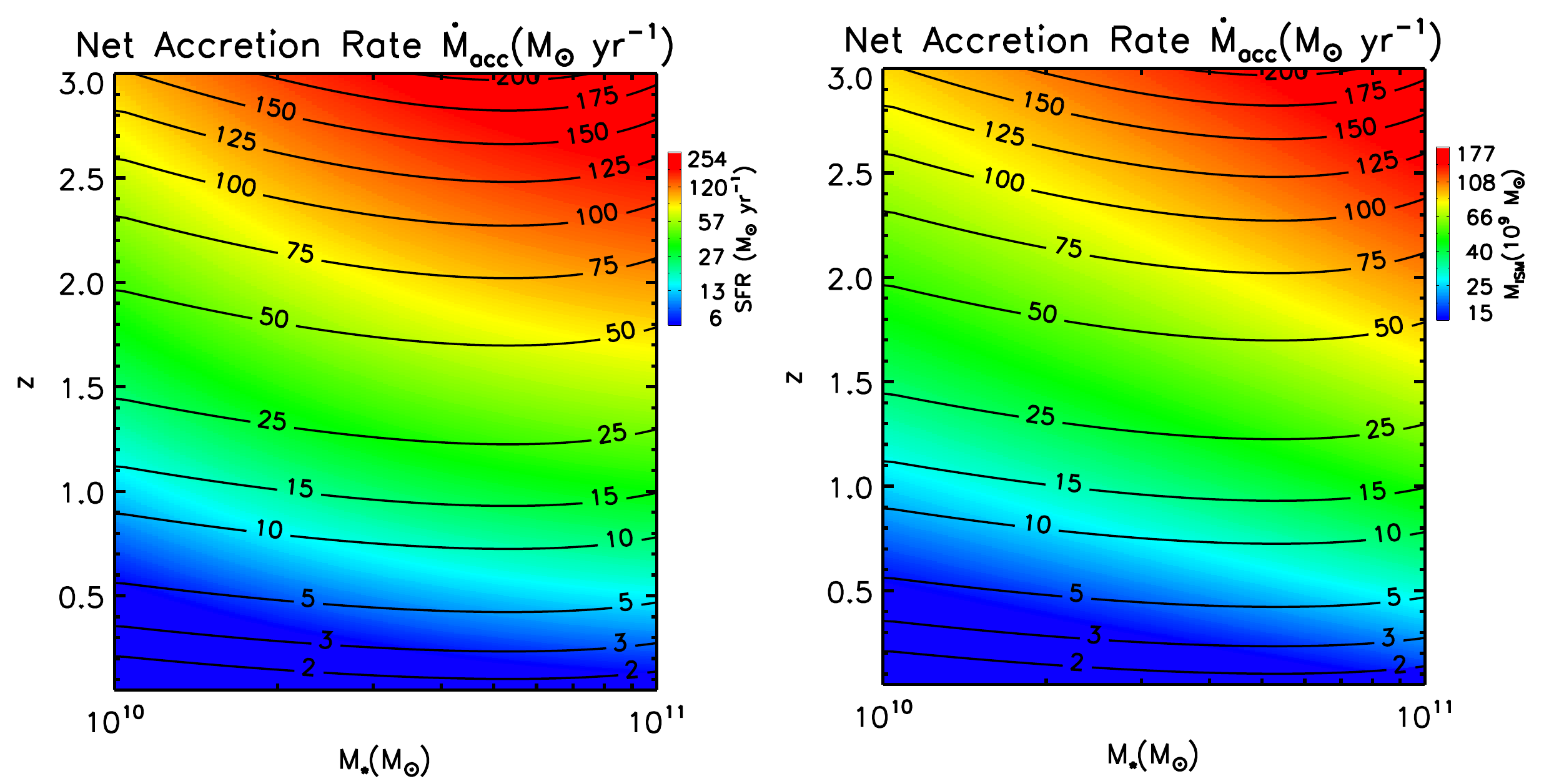}
\caption{The net accretion rates (contours) calculated using Equation \ref{mdot_eq} with the ISM masses given by Equation \ref{ism_fit}, the SFRs from Equation \ref{sfr_fit} and 
the MS tracking from the assumption of continuity in the evolution of the MS galaxy population. In the Left panel, the color background is SFR$_{MS}$; in the Right panel it is $M_{\rm ISM}$ on the MS in units of $10^9$\msun. We adopt a 30\% stellar mass-loss percentage \citep{lei11}.}
\label{accretion_rate_fig} 
\end{figure*}

 \begin{figure}[ht]
\epsscale{1.}  
\plotone{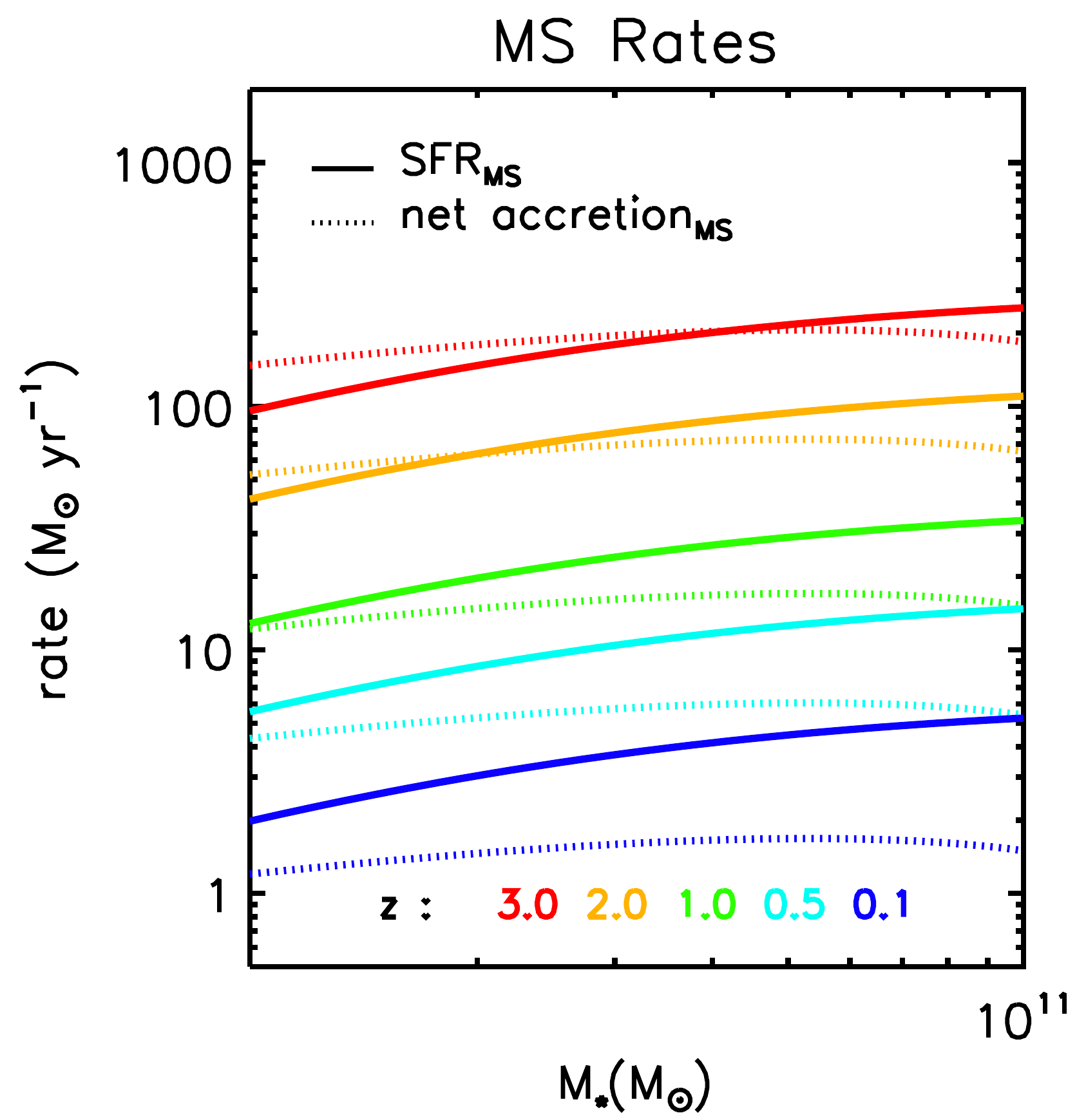}
\caption{The evolution of the SFR and the net accretion rate as a function of M$_{\rm *}$ and z. }
\label{all_evolution} 
\end{figure}

\section{Accretion Rates}\label{accretion}

Using the MS Continuity Principle (Section \ref{continuity}) and the ISM contents obtained from Equation \ref{ism_fit} with $\rm sSFR/sSFR_{MS} = 1$, we now derive the net accretion rates of MS galaxies required to maintain the MS evolutionary tracks. Along each evolutionary track (curved lines in Figure \ref{ms_evolution}), the rate balance must be given by:
\begin{eqnarray}  
{\rm dM_{\rm ISM} \over{dt}} = \rm  \dot{\rm M}_{acc} - (1-\rm f_{mass~return}) \times SFR, \label{mdot_eq}
 \end{eqnarray} 

\noindent assuming that major merging events are rare. $\rm f_{mass~return}$ is the fraction of stellar mass returned to the ISM through stellar mass-loss, taken to be 0.3 here \citep{lei11}. Since these paths are following the galaxies in a Lagrangian fashion, the time derivatives of a mass component M must be taken along the evolutionary track and
\begin{eqnarray}  
{\rm d M\over{dt}} &=& {\rm d M\over{dz}} {\rm dz \over{dt}} + {\rm d M\over{dM_{\rm *}}} {\rm dM_{\rm *} \over{dt}} \nonumber \\
{\rm d M\over{dt}} &=& {\rm d M\over{dz}} {\rm dz \over{dt}} + \rm SFR {\rm d M\over{dM_{\rm *}}}  \label{continuity_equation}.
 \end{eqnarray} 
 
  Figure \ref{accretion_rate_fig} shows the accretion rates using Equation \ref{mdot_eq}. The data used for this figure were generated from Equations \ref{ism_fit} and \ref{sfr_fit}, and thus are consistent with those earlier equations. These rates are $\sim100$ \msun yr$^{-1}$ at z $\sim$ 2.5.
 
 The power-law fit to the accretion rates is  
  \begin{eqnarray}  
 \dot{\rm M}_{acc} &=& \rm 2.27\pm0.24~\msun~yr^{-1}  \times   \rm (1+z)^{3.60\pm0.26} \times  \nonumber \\ 
&&\rm  \left( M_{10} ^{0.56\pm0.04}  -0.56\pm0.04 \times M_{10} ^{0.74\pm0.05}\right) \label{acc_fit} 
 \end{eqnarray} 
 
\noindent where $M_{10} = M_* / 10^{10} \msun$. The combination of two separate mass terms is required to match the curvatures shown in Figure \ref{accretion_rate_fig}; they don't have an obvious physical justification. The first term dominates at low mass and the second at higher masses. 


The value of the accretion rate 
 at z = 2 and M$_{\rm *} = 5\times10^{10}$ \msun~ is 73 \msun~yr$^{-1}$ with a $(1+z)^{\sim3.59}$ dependence (see Table \ref{equations2}). Theoretical estimates for the halo accretion rate of baryons yield lower values at z $\sim 2$ of $33$ \msun yr$^{-1}$ 
 with a (1+z)$^{2.5}$ dependence \citep{dek13}, that is a somewhat slower redshift evolution.

 Two important points to emphasize are: 1) these accretion rates should be viewed as \emph{net rates} (that is the accretion from the halos minus 
 any outflow rate from SF or AGN feedback) and 2) these rates 
 refer only to the MS galaxies where the evolutionary continuity is a valid assumption. 
  
 The derived accretion rates are required in order to maintain the SF in the early universe galaxies. Even though the existing gas 
 contents are enormous compared to present day galaxies, the observed SFRs will deplete this gas within $\sim5\times10^8$ yrs; this is short compared to the MS evolutionary timescales. The large accretion rates, comparable to the SFRs, 
 suggest that the higher SF efficiencies deduced for the high-z galaxies and for those above the MS may be dynamically driven by the infalling gas and mergers.  These processes will shock compress the galaxy disk gas and possibly be the cause of the higher SF efficiencies.

  It is worth noting that although one might think that the accretion rates could have been readily obtained simply from the evolution of 
the MS SFRs, this is not the case. One needs the mapping of $\rm M_{\rm ISM}$ and its change with time in order to estimate the first term on the 
right of Equation \ref{mdot_eq}. Figure \ref{all_evolution} shows the relative evolution of each of the major rate functions 
over cosmic time and stellar mass. Comparing the proximity of the curves as a function of redshift, one can see modest 
differential change in the accretion rates and SFRs  as a function of redshift and stellar mass. 

 \begin{figure}[ht]
\epsscale{1.}  
\plotone{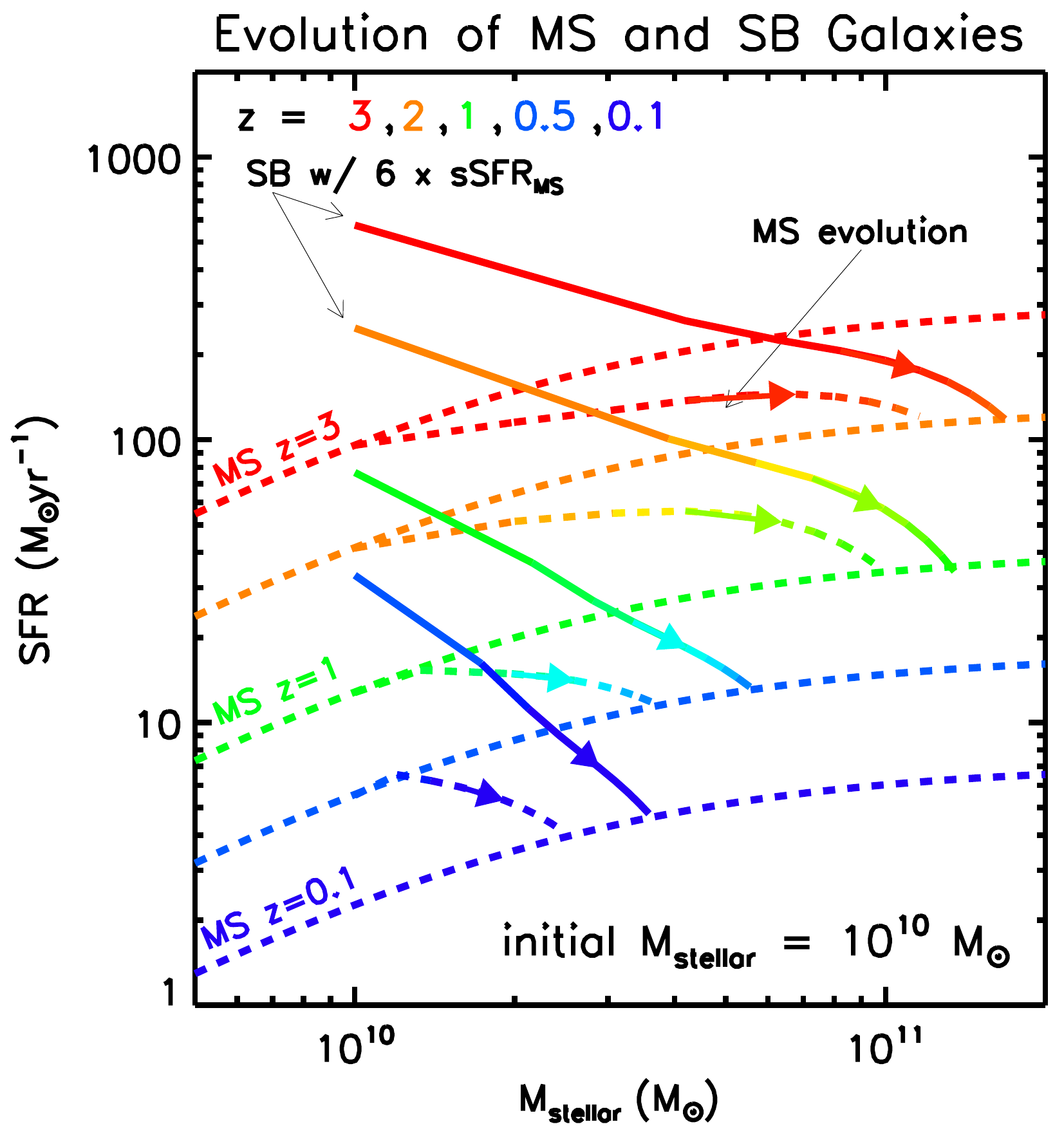}
\caption{Starburst galaxy evolution tracks (solid lines) compared with MS evolutionary tracks (dashed lines) for an initial $M_{\rm *} = 10^{10}$ \msun. 
The evolution is tracked between adjacent redshifts z = 3, 2, 1, 0.5 and 0.1. The SB galaxies start with an initial SFR 6 times larger than on the MS at the same mass and redshift.
Based on the derived fit in Equation \ref{ism_fit}, the SB will start with initial ISM content $6^{0.32} = 1.77$ times larger than the MS galaxy.}
\label{MS_SB_evolution} 
\end{figure}

\begin{figure*}[ht]
\epsscale{1.}  
\plotone{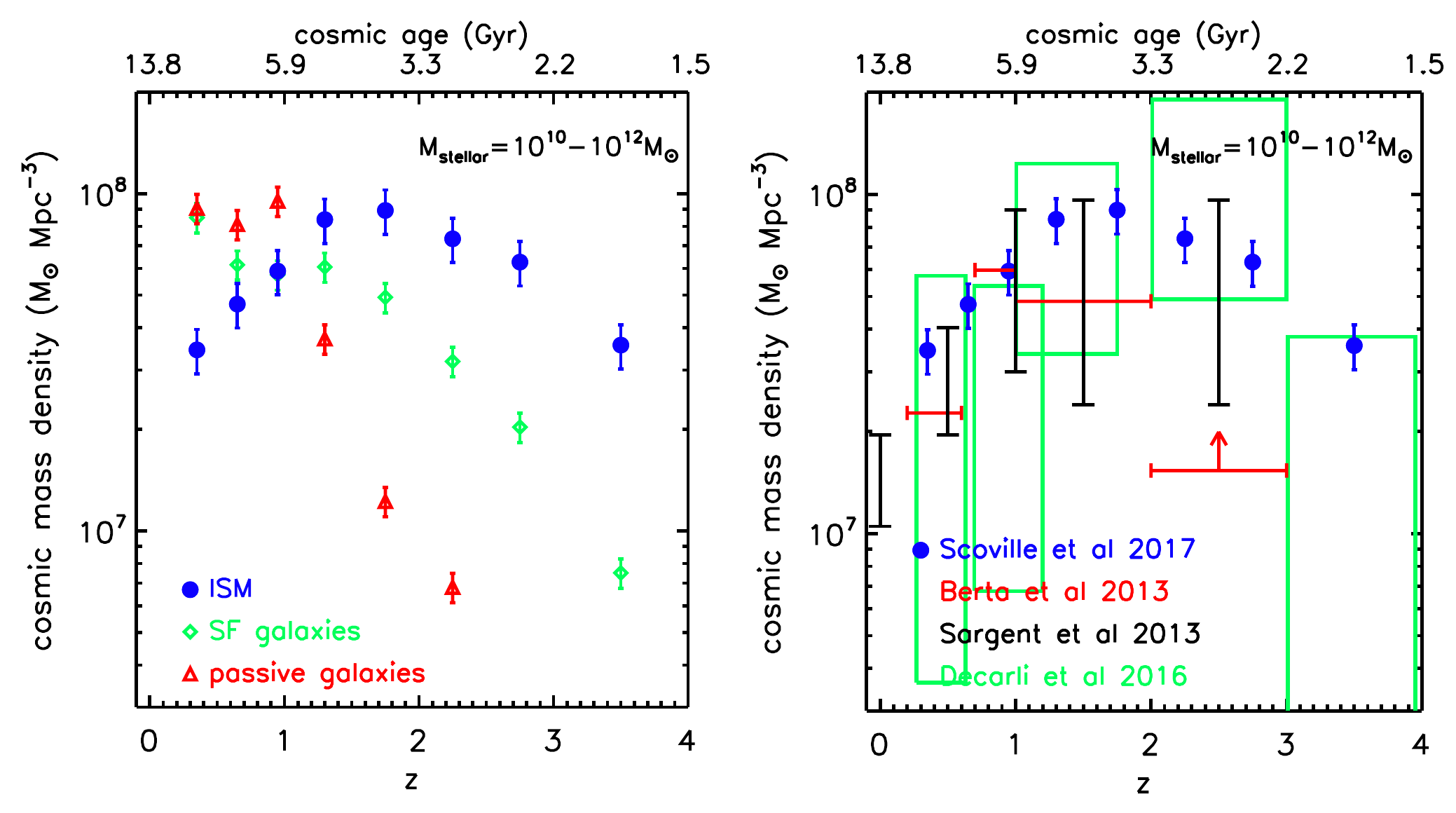}
\caption{Left: The cosmic evolution of ISM and stellar mass densities in the universe are shown for galaxies with stellar masses $\rm M_{\rm *} = 10^{10}$ to $10^{12}$ \msun.
 The galaxy stellar mass functions from \cite{ilb13} were used to calculate the ISM masses using Equation \ref{ism_fit}. Uncertainties in the stellar mass densities are typically $\pm10$\% for this range of $\rm M_{\rm *}$ \citep[see][Figure 8]{ilb13}; uncertainties in the ISM mass density also include an uncertainty of $\pm10$\% in the ISM masses when averaged over the population. (This does not include uncertainty in the calibration of the dust-based mass estimations.) Right: The ISM evolution derived here (blue points) is compared with that obtained by \cite{ber13} (red horizontal bars and a lower limit), the ALMA CO survey of \cite{dec16} (green boxs) with that derived by theoretical simulation \citep[black vertical error bars, ][]{sar13}}.
\label{lilly_madau}  
\end{figure*}

\begin{figure}[ht]
\epsscale{1.}  
\plotone{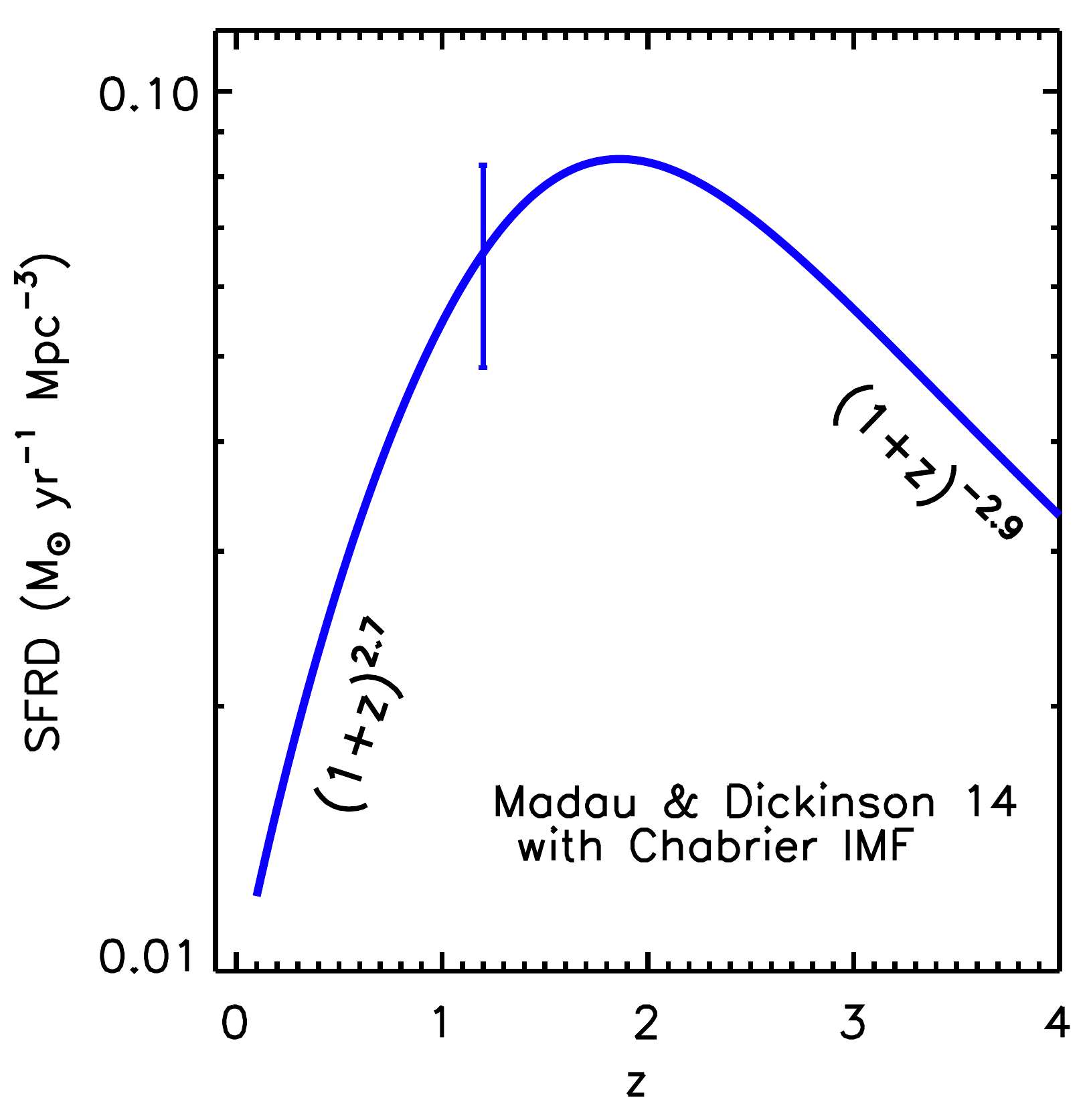}
\caption{The evolution of the cosmic star formation rate density (SFRD) from \cite{mad14} is shown for comparison with the overall evolution of ISM content in Figure \ref{lilly_madau}-Left. 
A typical uncertainty is shown as 0.1 dex \citep[see ][]{mad14}
(The relative scaling of z and 
the SFRD is the same as that in Figure \ref{lilly_madau}.) }
\label{mad_dickinson} 
\end{figure}

\begin{figure}[ht]
\epsscale{1.}  
\plotone{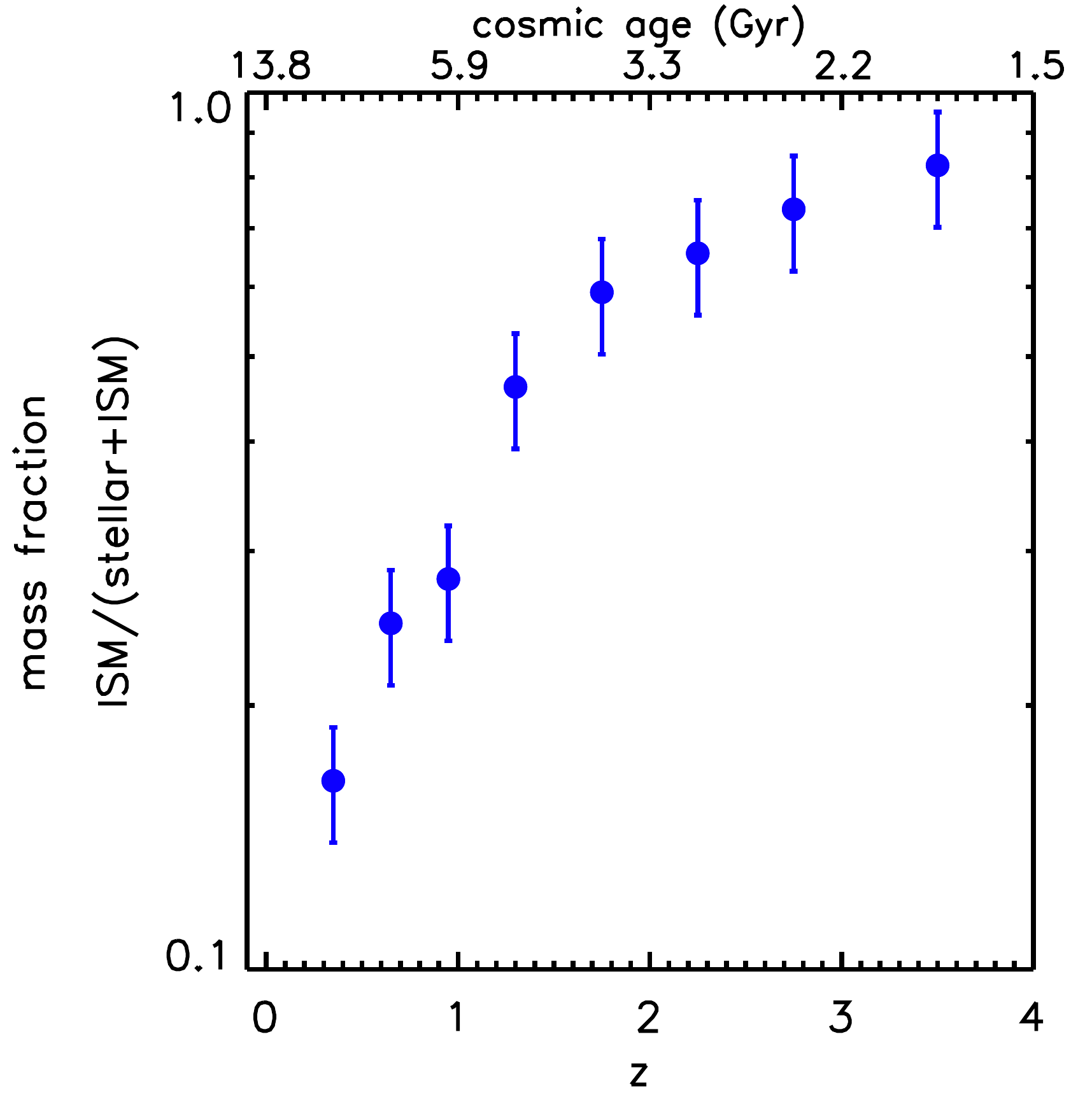}
\caption{The mass fraction of ISM are shown for galaxies with stellar masses $\rm M_{\rm *} = 10^{10}$ to $10^{12}$ \msun.}
\label{lilly_madau_2}  
\end{figure}

\section{SB versus MS Galaxy Evolution}\label{sb_ms}
 
 One might ask what are reasonable accretion rates to adopt for the SB galaxies above the MS, since they are not 
 necessarily obeying the continuity assumption? Here there appear two reasonable possibilities: either the accretion rate is similar to
 the MS galaxy at the same stellar mass, or if the elevation above the MS was a consequence of galactic merging, one might assume a rate 
 equal to twice that of an individual galaxy with half the stellar mass. The latter assumption would imply $\sqrt{\rm 2}$ higher accretion rate, thus 
 being consistent with the higher ISM masses of the galaxies above the MS. In this case, the higher SFRs will be maintained longer 
 than the simple depletion time it takes to reduce the pre-existing ISM mass back down to that of a MS galaxy. The same $\sqrt{\rm 2}$ factor of increase in the ISM mass 
 will arise from the merging of the pre-existing ISM masses of two galaxies of half the observed mass. This follows from the dependence of M$_{\rm ISM}$ on stellar mass, varying only as M$_{\rm *}^{0.30}$, rather than linearly. \emph {Thus, for two reasons, the notion that the SB region galaxies 
 are the result of major merging becomes quite attractive. }

 In Figure \ref{MS_SB_evolution}, we show the evolution of the SB galaxies with initial M$_{\rm *} = 10^{10}$ \msun and initial SFR a factor 6 above the MS. For reference, the evolution of MS galaxies with the same stellar 
 mass are also shown. For the SB galaxy, the initial ISM mass is $6^{0.32} = 1.77$ higher than the MS galaxy (see Equation \ref{ism_fit}).
 The accretion rate is taken as equal to that of a MS galaxy with the same $\rm M_{\rm *}$. This calculation shows that ultimately the SB ends up with approximately a factor 2 greater stellar mass -- due to 
 the larger initial ISM mass and the fact that the SB evolves more rapidly to higher stellar mass and thus accretes at a greater rate at high-z.

\section{Cosmic Evolution of ISM and Stellar Mass}\label{cosmic}

Using the mass functions (MF) of SF and passive galaxies \citep{ilb13}, we estimate the total cosmic mass density of ISM as a function of redshift using 
Equation \ref{ism_fit}. (This is the equivalent of the Lilly-Madau plot for the SFR density as a function of redshift.) We do this for the redshift range z = 0 to 4
and $M_{\rm *} = 10^{10}$ to $10^{12}$ \msun, a modest extrapolation of the ranges covered in the data presented here. Figure \ref{lilly_madau}-Left shows 
the derived cosmic mass densities of stars (SF and passive galaxies) and ISM as a function of redshift.  We applied Equation \ref{ism_fit} only to the SF galaxies and did not include any contribution from the passive galaxy population; to include the SB population we multiplied the ISM mass of the normal SF population by a factor of 1.1.  If the galaxy distribution were integrated down to stellar mass equal to $10^9$ \msun, 
the stellar (and presumably the ISM masses) are increased by 10 to 20\% \citep{ilb13}.  

The evolution of the ISM mass density shown in Figure \ref{lilly_madau}-Left is similar in magnitude to the theoretical predictions based on semi-analytic models by \cite{obr09,lag11,sar13} (see Figure 12 in \cite{car13}). However, all of their estimations exhibit a more constant density at z $>$ 1. The empirically based, prescriptive predictions of \cite{pop15} exhibit closer agreement with the evolution found by us; they predict a peak in the  
ISM gas at z $\simeq$ 1.8 and a falloff at higher and lower redshift similar to that seen in Figure \ref{lilly_madau}-Left.
\cite{ber13} estimated the 
evolution of ISM from the evolution of the SFRD by adopting a gas depletion time from \cite{tac13}. In Figure \ref{lilly_madau}-Right, their 
estimates are compared with ours and with the values from the simulation of \cite{sar13}.

The gas mass fractions computed for galaxies with $M_{\rm *} = 10^{10}$ to $10^{12}$ \msun~ are shown in Figure \ref{lilly_madau_2}. The ISM is dominant over the stellar mass down to z $\simeq 1.5$.  At z = 3 to 4 the gas mass fractions 
get up to $\sim 80$\% when averaged over the galaxy population. Thus, the evolution of ISM contents which peak at z $\simeq$ 2, is likely responsible for the peak in SF and AGN activity at that epoch \citep[see Figure \ref{mad_dickinson}][and references therein]{mad14}. At z = 4 down to 2, the buildup in the ISM density is almost identical to that of the cosmic SFRD shown in Figure \ref{lilly_madau}. (The ISM density point at z $\sim 0.3$ is uncertain since it relies on extrapolation of 
Equation \ref{ism_fit} to low M$_{\rm *}$ and low z, where there exist relatively few galaxies in our sample.)

\section{A Self-Consistency ~Test}\label{check}

Comparison of the MS evolutionary tracks shown in Figure \ref{MS_SB_evolution} with those shown in Figure \ref{ms_evolution} provides a critical confirmation 
of the self-consistency of our derived relations for the ISM masses (Equation \ref{ism_fit}, the SFR law (Equation \ref{sfr_fit}) and the accretion rates (Equation \ref{acc_fit}).
In Figure \ref{ms_evolution}, the MS evolution tracks were derived based solely on integration of the MS stellar mass and SFR relations as a function of redshift. Thus, those tracks make NO use of the relations derived here. 

On the other hand, the MS evolutionary tracks in Figure \ref{MS_SB_evolution} were obtained by integration in time of the galaxy evolution with the ISM mass, SFR and accretion given as a functions  of z, sSFR and M$_{\rm *}$ by Equations \ref{ism_fit}, \ref{sfr_fit}, and \ref{acc_fit}. Once the initial galaxy mass and z were specified, the initial ISM mass was taken from Equation \ref{ism_fit}. Then the SFR was calculated from Equation \ref{sfr_fit}, based on the ISM mass, and the accretion rate was calculated at each time step from Equation \ref{acc_fit}. The evolution was integrated in time using a 2nd order Runge Kutta method. The initial and final cosmic time for the integration was given by the redshifts for each pair of MS curves. Thus, the evolutionary 
tracks shown in  Figure \ref{MS_SB_evolution} rely entirely on the relations derived in this paper, specifically the dependence of ISM mass on MS location (determining the 
initial $M_*$ for the integration), and subsequently, the SFR as a function of ISM mass and the accretion rate as a function of stellar mass and redshift. 

Finding that the MS evolutionary tracks in Figure \ref{MS_SB_evolution} match those in Figure \ref{ms_evolution}, one concludes that the relations derived here are entirely consistent 
with the {\it a priori} known MS dependencies and the evolution of the MS. Although the tracks in Figure \ref{MS_SB_evolution} are shown in the figure only for initial M$_{\rm *} = 10^{10}$ \msun, we find a similar consistency for initial M$_{\rm *} = 5\times10^{10}$ \msun, indicating that the dependence on M$_{\rm *}$ in the equations is also self-consistent. 

\emph{In summary, the evolutionary tracks shown in Figure \ref{MS_SB_evolution} were derived using the 
 empirical fits to the ISM mass, the SFR as a function of ISM mass and the accretion rate function; the fact that the tracks for the MS galaxies follow closely those appearing in Figure \ref{ms_evolution} provides assurance the derived fits are consistent with the known MS evolution.}  [This does not provide a check on the validity of the relations above the MS,
 since we do not know {\it a priori} the end points of the SB evolutionary tracks.]

{\bf
\begin{deluxetable*}{lr}[h]

\large
\tablecaption{\bf{Summary of Relations for ISM, SFR and Accretion   }}

\tablehead{\colhead{} & \colhead{Eq. \#}   }
\startdata 
\\ 
 \boldmath $ \rm M_{\rm ISM} = \rm 7.07~\times 10^9~\msun    \times (1+z)^{~1.84}   \times (\rm sSFR/sSFR_{MS})^{0.32}   \times \left( {\rm M_{\rm *}\over{10^{10} \msun}}\right) 
^{0.30} $  & \ref{ism_fit}  \\
  
 \boldmath $ \rm SFR  = \rm 0.31~\msun~yr^{-1} \times  \left({M_{\rm ISM}\over{10^9 \msun}}\right)  \times   \left(1+z  \right) ^{1.04}  \times \left( \rm sSFR/sSFR_{MS}\right) ^{0.70}   \times \left( {\rm M_{\rm *}\over{10^{10} \msun}}\right)^{0.01} $  & \ref{sfr_fit} \\
 
 \boldmath $ \dot{\rm M}_{acc} = \rm 2.27~\msun~yr^{-1}  \times  \rm (1+z)^{3.60} \times \rm  \left( \left( {M_{\rm *}\over{10^{10} \msun}}\right) ^{0.56}  
-0.56 \times \left( {M_{\rm *}\over{10^{10} \msun}}\right) 
^{0.74}\right) $ & \ref{acc_fit}  \\
 
 \boldmath $ \rm  \tau_{dep}  = \rm 3.23 ~ Gyr~  \times  \rm 
(1+z)^{-1.04} \times (sSFR/sSFR_{MS})^{-0.70} 
\times  M_{10}^{-0.01} ~$ & \ref{dep_fit}  \\
\\ 
 \boldmath $ \rm  M_{ISM}/M_{\rm *}  = \rm 0.71 ~  \times \rm 
(1+z)^{1.84} \times  (sSFR/sSFR_{MS})^{0.32} 
\times  M_{10}^{-0.70} ~\rm $ & \ref{gas_m_fit}  \\
 \\
  \boldmath $ \rm  f_{gas} \equiv \rm {M_{\rm ISM} \over{M_{\rm *} + 
M_{\rm ISM}}} \nonumber =  \{ 1 ~+~1.41\times \rm 
(1+z)^{-1.84} \times \rm  (sSFR/sSFR_{MS})^{-0.32} \times  M_{10}^{0.70} \}^{-1} $   & \ref{fgas_fit} \\
 \\
 \\
 \enddata \label{equations1}
 \end{deluxetable*}

 \begin{deluxetable*}{lr}[h]

\large
\tablecaption{\bf{Relations Normalized to z = 2 and \boldmath $ \rm M_{*} =  5\times10^{10}$\msun} }

\tablehead{\colhead{} & \colhead{Eq. \#}   }
\startdata 

 \hline
  \hline
 \\
 \boldmath $\rm  Normalized ~to~ z ~= ~2 ~ and  ~ \rm M_{*} =  5\times10^{10}$\msun   \\
 
  \\ 
 \boldmath $\rm M_{\rm ISM}  = \rm 8.65\times 10^{10}~\msun~\times \left[(1+z)_2^{1.84} \times (sSFR/sSFR_{MS})^{0.32} \times  M_{*~5}^{0.30}\right]  $ & \ref{ism_fit}  \\
\\
 \boldmath $\rm SFR = \rm 9.9 \times  \left({M_{\rm ISM}\over{10^{10} \msun}}\right) ~\msun~yr^{-1}~~\times  \left[(1+z)_2^{1.04} \times  (sSFR/sSFR_{MS})^{0.70} \times  M_{*~5} ^{0.01} \right]$ & \ref{sfr_fit} \\
  \boldmath $\rm ~~~~  ~~~ = \rm 85   ~\msun~yr^{-1}~~\times  \left[\left({M_{\rm ISM}\over{8.65\times10^{10} \msun}}\right) \times (1+z)_2^{1.04} \times  (sSFR/sSFR_{MS})^{0.70} \times  M_{*~5} ^{0.01} \right]$ & \ref{sfr_fit} \\
\\
\boldmath  $ \dot{\rm M}_{acc} = \rm 73~\msun~yr^{-1}  \times   \left[2.3\times (1+z)_2^{3.60} \times  \left( M_{*~5}^{0.56}  -0.56 \times M_{*~5} ^{0.74}\right)\right] $ & \ref{acc_fit} 
  \\ 
  \\
\boldmath $ \rm \tau_{dep} \equiv \rm {M_{\rm ISM} \over{SFR}} = \rm 1.01 ~ Gyr~\times \left[ \rm (1+z)_2^{-1.04} \times  (sSFR/sSFR_{MS})^{-0.70} \times  M_{*~5}^{-0.01}\right]$     & \ref{dep_fit} \\
\\
\boldmath $ \rm gas/stellar \equiv \rm {M_{\rm ISM} \over{M_{\rm *}}} = \rm 1.74~\times  \left[\rm (1+z)_2^{1.84} \times  (sSFR/sSFR_{MS})^{0.32} \times  M_{*~5}^{-0.70}\right]$     & \ref{gas_m_fit} \\
\\
\boldmath $ \rm f_{gas} \equiv \rm {M_{\rm ISM} \over{ M_{\rm *} + M_{\rm ISM} }} =   0.63~\times \left[1.58/ \left(1 ~+~0.58\times \rm (1+z)_2^{-1.84} \times  (sSFR/sSFR_{MS})^{-0.32} \times  M_{*~5}^{0.70} \right)\right]$ & \ref{fgas_fit} \\
\\

\enddata \label{equations2}
\tablecomments{{\bf The equations are written in a form such that the quantity in \boldmath $\left[~\right]$ in each equation is equal to unity at z = 2 and \boldmath $M_{*} = 5\times10^{10}$\msun }. $\rm (1+z)_2$ is $(1+z)$ normalized to its value at z = 2 where (1+z)=3. M$_{*~5}$ is the stellar mass normalized to $5\times10^{10}$\msun.  As noted in Section \ref{relations}, the fourth relation is obtained by canceling out the M$_{\rm ISM}$ term 
in the second equation, not by division of the first equation by the second. See original equations in text for the uncertainties in the coefficients. }
\end{deluxetable*}
}

 \section{Comparison with Previous Work}\label{previous}

There are  now over 200 detections of CO line emission at high redshift (z $>$ 2) and we here compare our results with those studies. A couple caveats or cautions are
required when comparing these results: 1) Most of the
high-z CO detections are of CO (2-1) and (3-2); the inference of a molecular gas mass therefore requires an assumption for the scaling for the luminosity 
in these high J CO lines relative to the mass-calibrated CO (1-0) line, and 2) many of those studies have adopted a Galactic $\alpha_{\rm CO(1-0)}  = 2\times10^{20} \rm cm^{-2} (K~ \kms)^{-1}$, based largely on Galactic gamma ray observations, see review by \cite{bol13}. Here we have used $\alpha_{\rm CO(1-0)}  = 3\times10^{20} \rm cm^{-2} (\rm K~ \kms)^{-1}$ in our calibration of the dust emission to gas masses (see Appendix \ref{dust_app}). This $\alpha_{\rm CO(1-0)}$ is derived from correlation of the CO line luminosities and virial masses for resolved Galactic Giant Molecular Clouds (GMCs). We believe the former value is not as reliable -- it entails an 
assumption that the cosmic rays which produce the $\sim2$ MeV gamma rays by interaction with the gas fully penetrate the GMCs and even more suspect, that 
their density is constant with Galactic radius (see Appendix \ref{dust_app} and Appendix in \cite{sco16}). Below, we note where the works have used a 
different $\alpha_{CO}$ than that used in our calibration of the dust-based gas masses. 

The most extensive study of high- z CO (3-2) emission is that of \cite{tac13} who detected 38 galaxies at z $\sim 1.2$ and 14 at $z \sim 2.2$. (A somewhat more extensive sample including low redshift galaxies is included in \cite{gen15}). In their analysis they make use of 6 CO (2-1) detections at z $\sim 1.5$ \citep{dad10} and 6 at z $\sim 1$ \citep{mag12}. They find a single best-fit depletion time of 0.7$\pm0.2$ Gyr at z =1 to 2.2.
which would correspond to 1.1$\pm0.3$ Gyr for the CO conversion factor used here. Allowing for uncertainties in the assumptions used here and in the CO 
study, this is consistent with the depletion time estimated here -- $\simeq$1 Gyr  at z = 2 (see Table \ref{equations2}). 

These estimates can be compared with the z $\sim 0$
estimate of 1.24$\pm0.06$ Gyr \citep{sai11}, including HI. \cite{gen15} find that the depletion times vary as $(1+z)^{-0.3} \times (sSFR/sSFR_{MS})^{-0.5} \times M_*^{\sim0}$, compared with 
$(1+z)^{-1.04} \times (sSFR/sSFR_{MS})^{-0.70} \times M_*^{-0.01}$ from our work. Thus, we are finding 
a considerably steeper dependence on redshift, but similar dependencies on the elevation above the MS and the stellar mass (Table \ref{equations2}). They also see the gas contents 
varying linearly with the evolution of sSFR of the MS, whereas we find a 0.63 power-law dependence.
  
The molecular gas fractions (f$_{\rm gas} = \rm M_{\rm gas}/(M_{\rm gas}+M_{\rm *})$) found by \cite{tac13} were 0.49 and 0.47 at z = 1.2 and 2.2, respectively.  At z $\sim 2$, we find f$_{\rm gas} \simeq 0.6$ (Figure \ref{lilly_madau_2}) and 0.5 at z =1 for MS galaxies. \cite{tac13} used a lower $\alpha_{CO}$ 
so their gas fractions are necessarily lower than our estimates; however we see stronger evolution with redshift. Both the CO and dust-based estimates 
show a decreasing gas mass fraction at higher M$_{\rm *}$ (seen also by \cite{mag12}).
 \cite{gen15} find M$_{H2}/M_* \propto (1+z)^{3}$ whereas we find M$\rm_{H2}/M_* \propto \rm (1+z)^{1.84} \times sSFR/sSFR_{\rm MS})^{0.32} \times M_*^{-0.70}$ (Table \ref{equations2}); we thus have a 
more gradual evolution with redshift in the gas mass fractions. Some of the difference could be understood if the sampling in sSFR and $\rm M_*$ were very different between the 
two samples.

\cite{pop15} predict a decreasing gas-to-stellar mass ratio for higher stellar mass galaxies similar to that in Equation \ref{ism_fit}. Their work is based on 
tracking the halo gas contents and then using empirical relations from low z galaxies to model the galaxy properties such as size, SFR and H$_2$ content.  Figure 6 in \cite{pop15} indicates
M$\rm_{H2}/M_* \simeq 0.3$ and 1 at z = 1 and 3, averaged over $M_* = 10^{10} - 10^{12}$ \msun; Figure \ref{lilly_madau_2} indicates corresponding values of 0.42 and 3.5 after converting from M$\rm_{ISM}/(M_*+M_{ISM})$. Thus, at z = 1 there is reasonable agreement but at higher z we are finding larger gas mass fractions than \cite{pop15}. 

\cite{mag12} also examined the variations above the MS, obtaining  M$_{\rm gas} / \rm M_{\rm *} = 2.05\pm0.32 \times (\rm sSFR/sSFR_{\rm MS})^{0.87}$. This can be compared with our 
relation in Table \ref{equations2}, M$_{\rm gas} / \rm M_{\rm *} = 1.69\pm0.1 \times (\rm sSFR/sSFR_{\rm MS})^{0.32}$ at z = 2. Their work also uses the dust continuum to estimate masses. However, they also fit the far-infrared SED to obtain a variable $T_D$ -- which we have argued against on physical grounds (see Appendix \ref{dust_app}). \cite{sch16} observed a sample of 86 galaxies at 240 GHz with ALMA and detected 45 at z = 2.8 to 3.6. For the detected objects they obtain 
a median $\tau_{dep} = 0.68$ Gyr. Presumably this value would be larger if the non-detected sources were included. Their median gas-to-stellar mass ratio was $1.65\pm0.17$ for the 
45 objects. [Their ALMA dataset which is now public was included in the work presented here.]

In summary, there appears to be reasonable agreement between the results derived from CO line studies and those derived here, based on the RJ dust emission. 
This is indeed quite reassuring given the uncertainties in the higher J CO line and dust continuum calibrations. 

The very large galaxy sample presented here, based on the 
relatively short dust continuum measurements ($\sim 2$ min per galaxy with ALMA) and with uniform classification of the individual galaxy properties (redshift, stellar mass and SFR relative to the MS), have allowed us to thoroughly explore the variations of gas content with these properties.

\section{Summary}\label{summary}

We have analyzed the ISM gas contents of a sample of 708 galaxies at z $> 0.3$, having stellar masses determined from optical/NIR 
SED fitting and SFRs well-constrained since all are detected in the far-infrared with Herschel. We quantify the evolution of the ISM contents
as a function of redshift, M$_{\rm *}$ and sSFR (relative to the MS) by fitting simple power-law expressions to the observed M$_{\rm ISM}$ for a sample of 575 galaxies at z = 0.3 to 3.  
The fit for the ISM contents is then combined with the observed SFRs to capture the changing efficiencies of SFR per unit gas mass with redshift,
stellar mass and sSFR relative to the MS. The redshift evolutionary dependent term in the power laws was taken to be that of the SFRs on the MS; hence, the power-law 
fits readily show the relative evolution of the ISM and the efficiencies for SF, compared to the well-known evolution of the MS SFRs. The derived Equations are collected in Table \ref{equations2}
with normalization to z = 2 and M$_{\rm *} = 5\times10^{10}$ \msun.

\medskip
We find:
\begin{enumerate}
\item The ISM contents of SF galaxies, both MS and SB, increase to high redshift less rapidly than the SFRs (0.63 power of the SFR evolution function, ($SFR_{MS} \propto (1+z)^{2.9}$). 
\item Similarly, the ISM contents increase as the 0.36 power of the sSFR above the MS. 
\item The efficiency for forming stars per unit gas mass increases as the $\sim$0.32 power of redshift evolution of the SFRs and the 0.7 power of the elevation above the MS. .
\item Combining \# 1 to 3, it is clear that the increases in SF at high redshift and above the MS are due to both increased ISM masses and increased 
efficiency for converting gas to stars.
\item The enhanced SF activity of galaxies above the MS, due to both their larger gas masses and higher SF efficiencies, suggests
very plausibly that these starburst galaxies are the result of galaxy merging. 
\item The ISM contents increase as $\rm M_{\rm *}^{0.3}$, implying that the gas mass fractions decrease 
at higher stellar masses. The SF efficiency (SFR per unit gas mass) is virtually independent of M$_{\rm *}$.
\item We then estimate the accretion rates under the reasonable assumption of continuity from one epoch to the next in the MS galaxy populations. The 
derived net accretion rates (required to maintain the SF activity) are extreme, exceeding 50\msun yr$^{-1}$ above z =2. The specific accretion rates (normalized to the stellar mass of the galaxy) decrease for higher mass galaxies. 
\item An analytic fit for the accretion rates shows that the accretion increases even more rapidly at high redshift than the MS SFRs. Thus, it is the 
evolution of this accretion which drives the galaxy evolution in the early universe. In fact, the higher SF efficiencies at high redshift and above the MS may be 
linked to dynamical compression of the ISM by the infalling gas and minor mergers. 
\item To illustrate the power of these results, we can now chart the evolutionary paths of galaxies (both MS and SB galaxies) over cosmic time as they grow in mass
and their SF dies out (due to decreased accretion) (see Figure \ref{MS_SB_evolution})
\end{enumerate}


\subsection{Comments and Implications}

The variations of ISM masses, accretion and their relation to star formation have been explored with the most extensive sample yet of 
high redshift galaxies. Although the deduced estimates are consistent with existing studies using the CO lines \citep{tac10,dad10,gen10,rie11,ivi11,mag12a,sai13,car13,tac13,bol15,gen15}, the sample of galaxies used here has the virtue that it maps the parameter 
space of z, $M_{\rm *}$, and sSFR quite effectively out to z = 3 using high quality and uniform ancillary data from the COSMOS survey field. We thus can simultaneously constrain the functional dependencies on redshift, sSFR relative to the MS and stellar mass. Our technique also does not suffer from the uncertainties introduced by 
variable excitation in the higher-J CO lines. 

There are obvious extensions which need to be done in this field: extending with larger samples at z $< 1$ and z $> 3$. The former is straightforward 
since the lower z galaxies can be observed on the RJ tail in ALMA band 7 (for which the ALMA sensitivity is excellent and the fluxes are high); 
the latter will be more time-consuming since the observations must shift to Band 6 or even lower frequencies to stay on the Rayleigh-Jeans tail and the fluxes will 
be lower. In addition, the number of z $>$ 3 galaxies with measured far-infrared luminosities is much less. 

Obtaining high quality CO (1-0) data on a subset of the galaxies analyzed here is also a high priority in order to firm up the calibration of dust flux to ISM masses 
at high redshift, and to determine the range of stellar masses above which it is valid. This will be both time-consuming and fraught with difficulties in the analysis 
since the CO line has its own calibration issues. But the spectral line data will also provide very helpful dynamical mass estimates -- these can certainly be a 
useful reality check. 

A major uncertainty for both the dust and the CO line studies is, of course, their dependence on metallicity (Z). Both are probably more robust than
is generally assumed in the community. The CO line is heavily saturated (even in Galactic GMCs which have typical $\rm \tau_{CO (1-0)} \gtrsim 10$). To underscore this point, we note that the $^{13}$CO emission 
line is typically $\sim1/5$ of the CO line flux in Galactic GMCs despite the much lower $^{13}$C/C abundance ($\sim$1/60 to 1/90). Thus, the line luminosity must scale 
as the $\sim$1/3 power of the CO abundance, and hence the metallicity. 

With respect to the dust emission as a probe of ISM, it is reassuring that the dust-to-gas abundance ratio in low redshift galaxies appears fairly constant at $\sim$1\% by mass from solar down to 1/5 solar metallicity \citep[see][]{dra07a} and \cite[][Figure 16]{ber16} (although why this is the
case is not understood).

Our finding that the ISM-to-stellar mass ratio and the accretion rates are both generally higher for lower mass galaxies has implications 
for the gas-phase metallicites of galaxies. Assuming that the metallicity of freshly accreting gas is significantly lower than that of the internal gas or the stars 
formed out of the gas, one would expect the gas phase metallicity to increase in higher stellar mass galaxies. This is, of course, known to be true; and it is
a major motivation for our restriction to galaxies with relatively high M$_{\rm *}$. It is also clear that a so-called `closed box' model for the evolution of metal content has little 
physical justification in light of the extremely large accretion rates derived here.

\acknowledgments

We thank Zara Scoville for proof reading the manuscript and we thank the referee for a number of constructive suggestions. In addition, several useful references on the Galactic GMCs were provided by John
Carpenter and John Bally and good suggestions were provided by Fabian Walter.
This paper makes use of the following ALMA data:
  ADS/JAO.ALMA 2011.0.00097.S, 2012.1.00076.S, 
  2012.1.00523.S, 2013.1.00034.S, 2013.1.00111.S, 2015.1.00137.S, 2013.1.00118.S, and 2013.1.00151.S. 
   We plan to release the images used here via the IPAC/IRSA COSMOS archive at http://irsa.ipac.caltech.edu/data/COSMOS/ in 2017. ALMA is a partnership of ESO (representing its member states), 
  NSF (USA) and NINS (Japan), together with NRC (Canada), NSC and ASIAA (Taiwan), and KASI 
  (Republic of Korea), in cooperation with the Republic of Chile. 
  The Joint ALMA Observatory is operated by ESO, AUI/NRAO and NAOJ.
			The National Radio Astronomy Observatory is a facility of the National
			Science Foundation operated under cooperative agreement by Associated
			Universities, Inc. RJI acknowledges support from the European Research Council (ERC) in the form of Advanced Grant, 321302, COSMICISM. ST acknowledges support from the European Research Council (ERC) in the form of Consolidator Grant, 648179, ConTExt. B.D. acknowledges financial support from NASA through the Astrophysics Data Analysis Program (ADAP), grant number NNX12AE20G.
\bibliography{scoville_dust}{}

 \appendix

\section{Long Wavelength Dust Continuum as an ISM Mass Tracer}\label{dust_app}

Here, we very briefly summarize the more thorough discussion in the Appendix in \cite{sco16} which establishes the foundation for using the long wavelength dust continuum 
as a tracer of ISM mass in high redshift galaxies.

 The empirical calibration of the technique \citep{sco16} is based on 3 different low and high redshift galaxy samples: 1) a sample of 30 local star forming galaxies; 2) 12 low-z Ultraluminous Infrared Galaxies (ULIRGs);  and 3) 30 z $\sim$ 2 submm galaxies (SMGs). These three samples with 72 galaxies are restricted to only those galaxies having good estimates of the 
total, source-integrated, long-wavelength continuum and complete mapping of CO (1-0). (We avoid using higher-J CO lines since only the 1-0 transition has been well-calibrated using large samples of 
Galactic GMCs with viral mass estimates; the higher CO lines have variable flux ratios with respect to the 1-0 line so they are unlikely to be as reliable in mass estimations.)
In calibrating the CO(1-0) masses, we have adopted $\alpha_{CO(1-0)}  = 3\times10^{20} cm^{-2} (K \kms)^{-1}$ which 
is derived from correlation of the CO line luminosities and virial masses for resolved Galactic GMCs. We believe this is more correct than the value obtained from Galactic gamma ray 
surveys ($\alpha = 2\times10^{20}$) \citep[see][]{bol13} since the latter requires the questionable assumptions: 1) that the cosmic rays which produce the $\sim2$ MeV gamma rays by interaction with the gas fully penetrate the GMCs and 2) that 
the cosmic ray density is constant with Galactic radius. Obviously if one adopts the latter value, our derived scaling for the dust-based ISM masses must be reduced by a 
factor of 2/3. (See the appendix in \citep{sco16} for a more extensive discussion.)

\begin{figure*}[ht]
\epsscale{1}  
\plotone{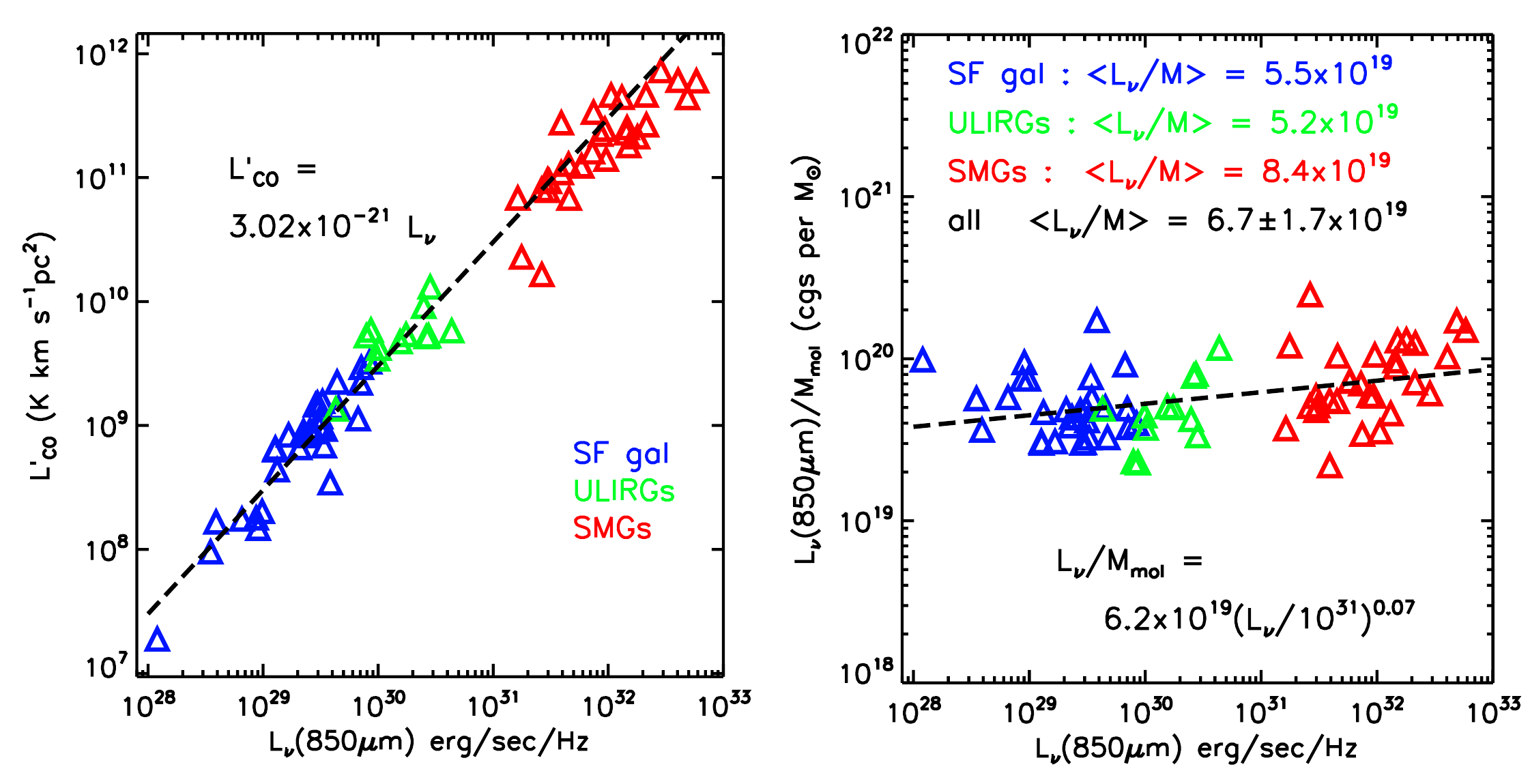}
\caption{{\bf Left:} The CO(1-0) luminosity and $L_{\nu}$ at 850$\mu$m are shown for three samples 
of galaxies -- normal low-z star forming galaxies, low-z ULIRGs and z $\sim$ 2 SMGs. All galaxies were selected to have global measurements of CO (1-0) and Rayleigh-Jeans dust continuum fluxes. The large range in apparent luminosities is enhanced by including the high redshift SMGs, many of which 
in this sample are strongly lensed. {\bf Right:} The ratio of $ L_{\nu}$ at 850$\mu$m to $M_{\rm ISM}$ is shown for the three samples of galaxies, indicating a very similar proportionality constant 
between the dust continuum flux and the molecular masses derived from CO(1-0) emission. The molecular masses were estimated from the CO (1-0) luminosities using a single standard Galactic  $X_{CO} = 3\times10^{20}$ N(H$_2$) cm$^{-2}$ (K km s$^{-1})^{-1}$. }
\label{empir_cal} 
\end{figure*}

Figure \ref{empir_cal} shows the ratio of specific luminosity at rest frame $\lambda = 850 \mu$m to that of the CO (1-0) line, and one clearly sees a quite similar ratio
of RJ dust continuum to CO luminosity. Using a standard Galactic CO (1-0) conversion factor, we then obtain the relation by which we convert the RJ dust continuum
to ISM masses:
  \begin{eqnarray}
M_{\rm ISM}   &=& 1.78 ~S_{\nu_{obs}}[\rm mJy]  ~ (1+z)^{-4.8} ~   \left({ \nu_{850\mu \rm m}\over{\nu_{obs} }}\right)^{3.8}   ({\it{d}_L \rm{[Gpc ]}})^{2} \nonumber \\
 && \times ~\left\{{6.7\times10^{19}\over{\alpha_{850}  }} \right\}    ~{\it{\Gamma_{0}} \over{\it{\Gamma_{RJ}}}}~   ~10^{10}\msun    ~. \label{mass_eq}  \label{dust_eq}
  \end{eqnarray}
  
In Equation  \ref{dust_eq}, $\Gamma$ is a correction for departures from strict $\nu^2$ of the RJ continuum, and $\alpha_{850} = 6.7\pm1.7\times 10^{19} \rm erg ~sec^{-1} Hz^{-1} {\msun}^{-1}$ is the derived calibration constant 
between 850$\mu$m luminosity and ISM mass. We have adopted a dust opacity spectral index $\beta = 1.8$, based on the determinations of the Planck observations 
in the Galaxy \citep{pla11a,pla11b}. [\cite{ber16} and \cite{bia13} provide extensive discussions of possible variations in $\beta$.]

\begin{figure}[ht]
\epsscale{0.5}  
\plotone{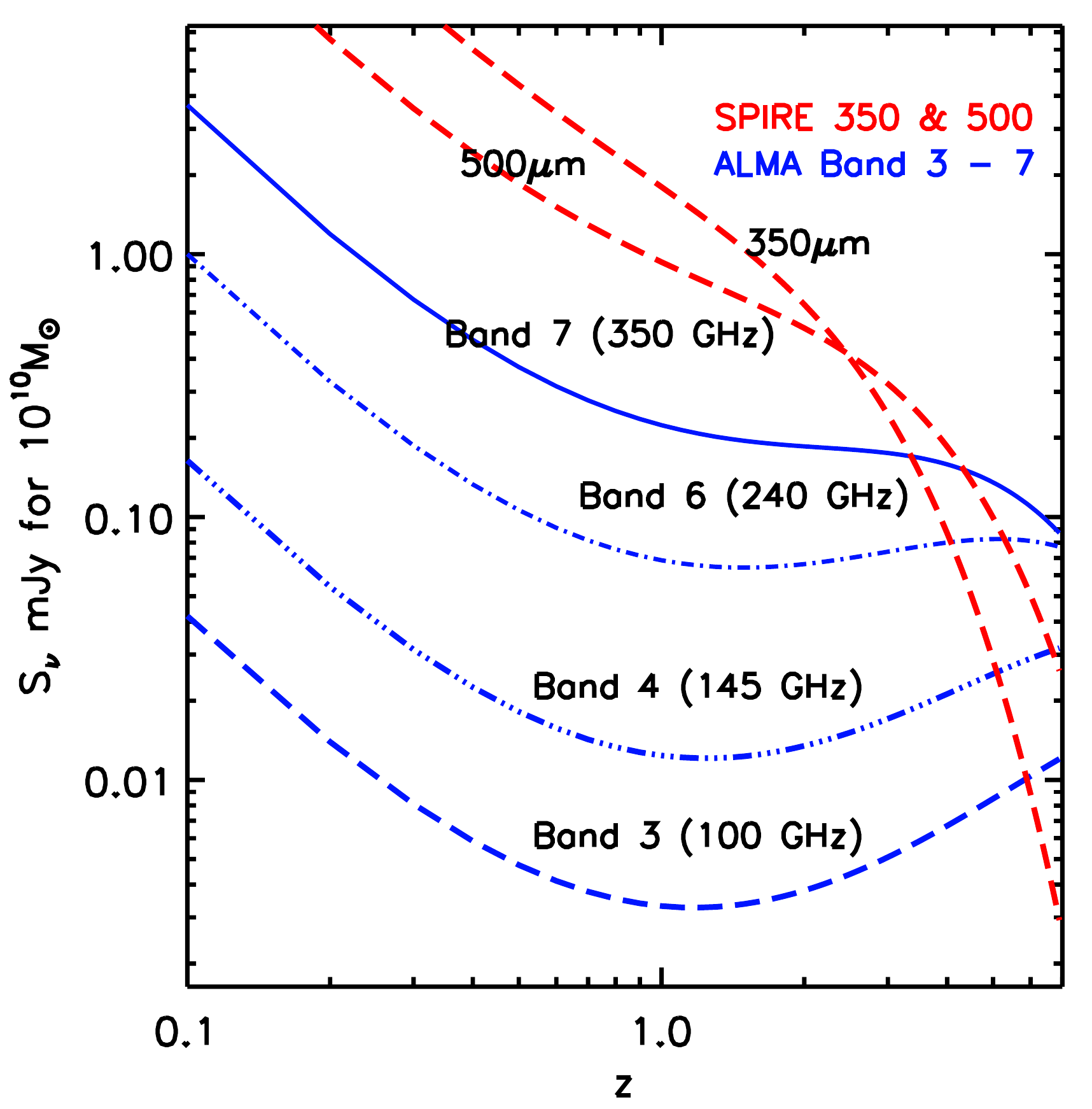}
\caption{The expected continuum fluxes for the ALMA bands at 100, 145, 240 and 350 GHz and for SPIRE 350 and 500$\mu$m for M$_{ISM} = 10^{10}$\msun ~derived using the empirical calibration $\alpha = 6.7\times10^{19}$ \citep{sco16}, 
an emissivity power law index $\beta = 1.8$ and including the RJ departure coefficient $\Gamma_{RJ} (25K)$. Since the point source flux sensitivities of ALMA 
in the 4 bands are quite similar, the optimum strategy is to use Band 7 out to z $\sim 2 -3$; above z = 3. Lower frequency ALMA bands are required to avoid large uncertainties in the RJ correction.}
\label{alma_obs} 
\end{figure}

 In the present work, the conversion of the fluxes measured with ALMA  to ISM masses is done with Equation \ref{dust_eq}. Using Equation \ref{dust_eq}, the predicted fluxes for a fiducial ISM mass of $10^{10}$\msun ~in the ALMA bands are shown in Figure \ref{alma_obs} as a function 
of redshift.

In the current work we restricted the observed galaxies to be relatively massive ($M_{*} > 10^{10}$ \msun) since they should have close to solar metallicity and presumably not low
dust-to-gas abundance ratios. We note that for the first factor $\sim5$ down from solar metallicity in the galaxies analyzed by \cite{dra07b}, there is virtually no variation in the 
dust-to-gas abundance when one considers only objects with complete mapping in both CO (1-0) and the dust continuum \citep[see][]{sco16}. In fact, the dust abundance is likely to be more 
robust than CO, which can suffer depletion due to UV photo-dissociation as the metallicity drops.

 \section{Dust Emission Spectral Index and Dust Temperature}\label{duals}

 \begin{figure}[ht]
\epsscale{0.75}  
\plotone{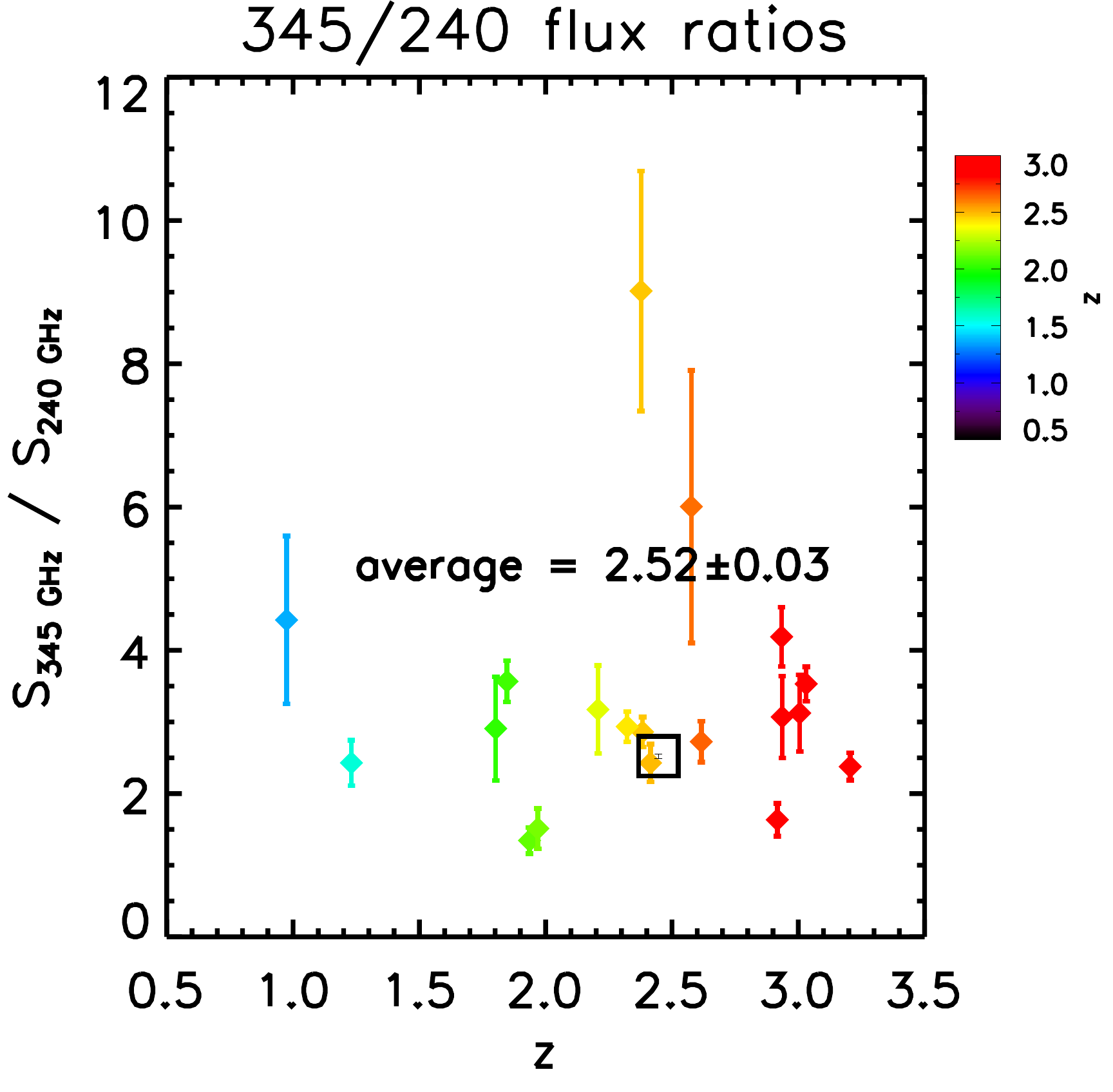}
\caption{A  summary of the 345 / 240 GHz flux ratios are shown for 19 sources having dual band measurements (see Table \ref{duplicates}). The weighted mean value (shown by the square box) is 2.52$\pm$0.03.The mean expected flux ratio for 25K dust temperature and opacity spectral index of 1.8 is 2.52 across the range of redshifts sampled above (see Figure 2 \cite{sco16}). }
\label{ratios} 
\end{figure}

The 19 objects which have both Band 6 and 7 measurements can be used to check for consistency with the spectral index of the submm dust emissivity ($\kappa_{\nu}$ = 1.8) and adopted dust
temperature ($T_D = 25$ K) used for translating fluxes to masses. The SNR-weighted mean value of 
the Band 7 / Band 6 flux ratio is $2.52\pm0.03$ -- consistent with the mean expected ratio of 2.52 (once one accounts for the departures 
from a strict Rayleigh-Jeans approximation). These measurements are shown in Figure \ref{ratios}.  We have not explored constraining the dust temperature variations based on the range of the two-band ratios.  

\begin{deluxetable*}{llcccccccccc}[h]
\footnotesize
\tablecaption{\bf{Sources with 240 and 345 GHz Measurements}  }

\tablehead{ 
 \colhead{RA} &  \colhead{Dec}  & \colhead{z} & \colhead{$<\nu_{240}>$} & \colhead{S$_{\nu}$}  &  \colhead{$<\nu_{345}>$} & \colhead{S$_{\nu}$}  & \colhead{ratio} \\ 
 \colhead{} & \colhead{} & \colhead{} & \colhead{GHz} & \colhead{mJy} & \colhead{GHz} & \colhead{mJy}  }
\startdata 
\\
 149.64888 &    2.59813 &  2.94 & 240.0 &  1.12 & 343.5 &  3.43 &  3.07$\pm$0.6  \\
 149.65558 &    2.71626 &  2.42 & 240.0 &  3.95 & 343.5 &  9.58 &  2.43$\pm$0.3  \\
 149.66782 &    2.08743 &  0.97 & 240.0 &  1.35 & 341.9 &  5.95 &  4.42$\pm$1.2  \\
 149.72572 &    2.27947 &  3.03 & 240.0 &  4.24 & 343.5 & 14.98 &  3.53$\pm$0.2  \\
 149.75085 &    1.85219 &  1.85 & 240.0 &  3.50 & 343.5 & 12.48 &  3.57$\pm$0.3  \\
 149.75095 &    1.85469 &  3.21 & 240.0 &  3.12 & 323.3 &  7.42 &  2.38$\pm$0.2  \\
 149.87177 &    2.21219 &  2.38 & 240.0 &  2.23 & 343.5 & 20.08 &  9.02$\pm$1.7  \\
 149.92032 &    2.02038 &  2.38 & 240.0 &  4.50 & 343.5 & 12.88 &  2.86$\pm$0.2  \\
 150.03714 &    2.66954 &  2.93 & 240.0 &  1.37 & 343.5 &  5.73 &  4.19$\pm$0.4  \\
 150.08232 &    2.53454 &  2.62 & 240.0 &  3.20 & 343.5 &  8.71 &  2.72$\pm$0.3  \\
 150.09865 &    2.36537 &  2.58 & 240.0 &  0.62 & 343.5 &  3.70 &  6.01$\pm$1.9  \\
 150.10616 &    2.05351 &  1.23 & 240.0 &  2.60 & 343.5 &  6.32 &  2.43$\pm$0.3  \\
 150.14705 &    2.73147 &  2.21 & 240.0 &  1.94 & 343.5 &  6.16 &  3.17$\pm$0.6  \\
 150.15022 &    2.47518 &  1.94 & 240.0 &  1.37 & 343.5 &  1.84 &  1.34$\pm$0.2  \\
 150.17995 &    2.08864 &  1.97 & 240.0 &  0.99 & 343.5 &  1.50 &  1.51$\pm$0.3  \\
 150.31131 &    2.58844 &  3.01 & 240.0 &  2.20 & 344.8 &  6.87 &  3.12$\pm$0.5  \\
 150.31342 &    2.71619 &  1.80 & 245.0 &  0.54 & 344.8 &  1.57 &  2.91$\pm$0.7  \\
 150.34572 &    2.33485 &  2.92 & 240.0 &  2.41 & 344.8 &  3.93 &  1.63$\pm$0.2  \\
 150.34946 &    1.93700 &  2.32 & 240.0 &  2.58 & 343.5 &  7.58 &  2.93$\pm$0.2  \\
  \\ 
\enddata\label{duplicates}
\tablecomments{The expected flux ratio for 25K dust temperature and opacity spectral index of 1.8 is $2.52\pm0.01$ across the range of redshifts sampled above. }
\end{deluxetable*}

\section{Fitting Co-variances}\label{mcmc}

The covariance distributions obtained from the MCMC fitting for Equations \ref{ism_fit} and \ref{sfr_fit} are shown. 
This is a Bayesian method for linear regression that takes into account measurement errors in all variables, as well as intrinsic scatter. A Markov chain Monte Carlo algorithm (Gibbs sampling) is used to randomly sample the posterior distribution. 

\begin{figure}[ht]
\epsscale{1.}  
\plottwo{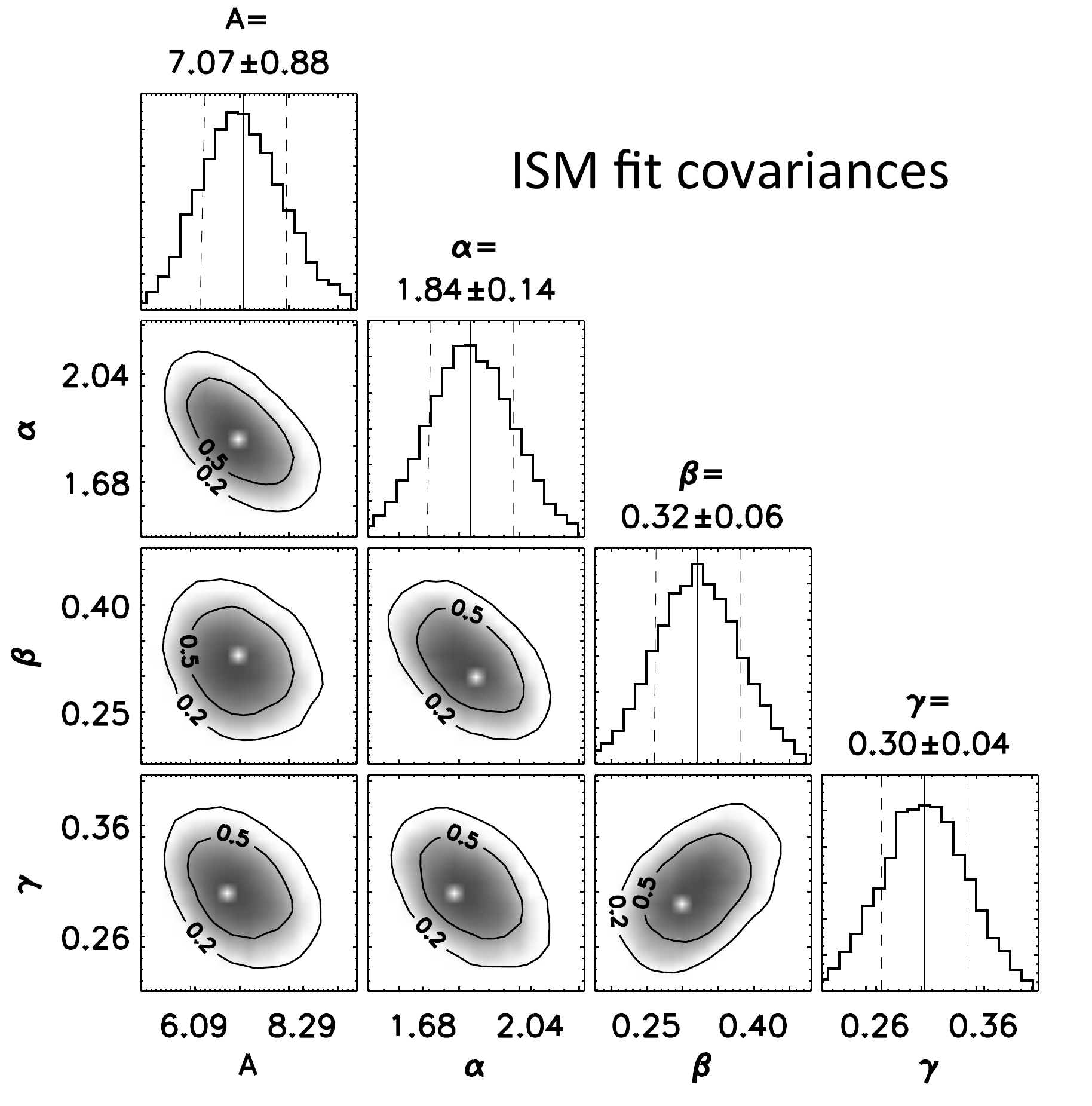}{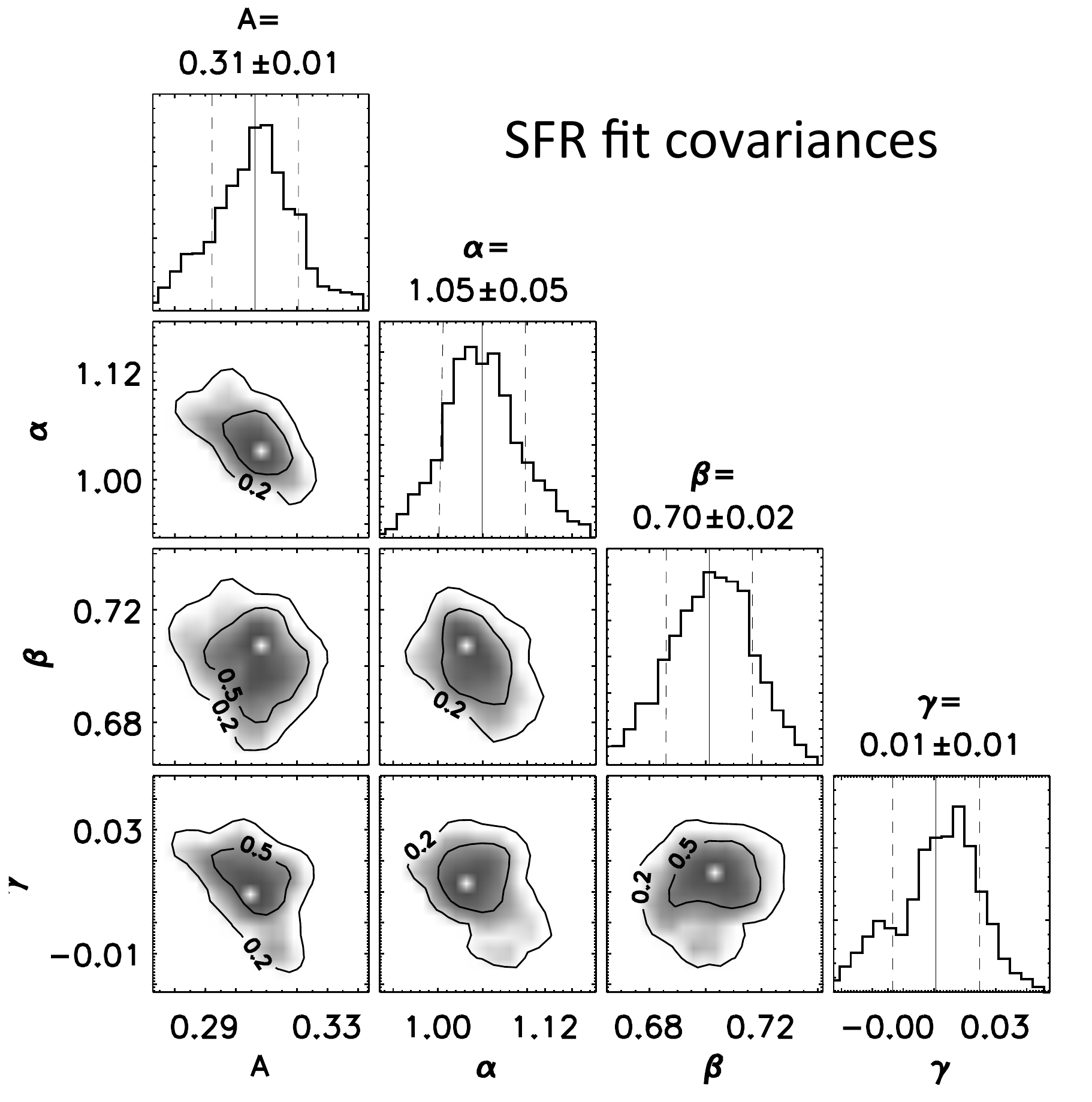}
\caption{The covariance distributions for the fits obtained in Equations \ref{ism_fit} and \ref{sfr_fit} are shown in the Left and Right panels, respectively. The parameters 
A, $\alpha$, $\beta$ and $\gamma$ correspond to the lead scale factor and the exponents in the Equations.}
\label{cov} 
\end{figure}

\end{document}